\documentclass[preprint,3p,10pt]{elsarticle}

\usepackage{amssymb}
\usepackage{hyperref}
\usepackage{xcolor}
\usepackage[export]{adjustbox}

\usepackage{amsthm}
\usepackage{amsmath}
\usepackage{tabularx}
\usepackage{tikz}

\journal{Journal of Fluids and Structures}

\begin{document}

\begin{frontmatter}



\title{Unsteady load alleviation on highly flexible bio-inspired\\wings in longitudinally oscillating freestreams}


\author[ISAE-SUPAERO]{{\'A}lvaro Mart{\'i}nez-S{\'a}nchez}
\author[ISAE-SUPAERO]{{\'A}lvaro Achirica-Villameriel}
\author[ISAE-SUPAERO]{Nicolas Dou{\'e}}
\author[ISAE-SUPAERO]{\\Val{\'e}rie Ferrand}
\author[ISAE-SUPAERO]{Erwin R. Gowree}

\affiliation[ISAE-SUPAERO]{organization={ISAE-SUPAERO, University of Toulouse},
            addressline={10 Avenue Edouard Belin}, 
            city={Toulouse},
            postcode={31400}, 
            country={France}}

\begin{abstract}

This study delves into the aerodynamic behavior of a highly flexible NACA 0012 aerofoil, drawing inspiration from avian feathers to handle a gust. We first examined unsteady flow on a rigid wing both experimentally and numerically and then explored the implications of introducing wing flexibility purely numerically. Our findings underscore the potential of composite materials in alleviating the oscillating aerodynamic forces on a wing under a streamwise gust. This behavior is attributed to its capacity to destabilize the laminar separation bubble, fostering a more stable turbulent boundary layer. While direct avian evidence remains limited, it is postulated that in nature, such mechanisms could mitigate undesired wing flapping, optimizing energy consumption in perturbed environments.

\end{abstract}




\end{frontmatter}


\section{Introduction}

Unmanned aerial systems (UAS) are increasingly being deployed in harsh environmental conditions for a broad array of applications, such as fire detection, search and rescue, urban air mobility, and wildlife monitoring, to name a few \cite{hassanalian2017, jimenez2019}. Such conditions consist of flow disturbances, including turbulent eddies or gusts emanating from urban canyons, building wakes, or landscape topography \cite{adkins2019}, which reduce the operability of the aerial vehicle. These systems typically have small dimensions, on the order of a metre, and operating speeds between 10 and 15 m/s, which are comparable to the gusts and eddies encountered in disturbed environments \cite{Watkins2006, Berry2012, Jones2022}. Due to their small size and low speed, these vehicles are more susceptible to flight stability and control issues arising from unstable flow environments. The inherently unsteady flow over low Reynolds number flyers as a result of laminar separation, recirculation bubbles, or boundary layer transition, and the associated fluid-structure interaction remain inadequately understood, with limited predictive capabilities and experimental evidence. Moreover, recent studies \cite{Jaroslawski2023a, Jaroslawski2023b} have demonstrated that the dynamics of the laminar separation bubble can be altered in the presence of the high levels of turbulence intensity typically encountered in urban environments. 
Therefore, further investigation into the behaviour of these systems in perturbed environments is essential for enhancing the existing numerical prediction methodologies used for their design and optimisation, as well as for developing control strategies, such as wing morphing. This is particularly crucial in densely populated areas where these systems must operate reliably and safely.

Coupled with the above, the harsher environmental conditions brought about by climate change will impose stricter limitations on the operability of aerial vehicles. Clear-air turbulence (CAT) is responsible for 24\% of aircraft accidents at high altitudes, and due to climate change, the frequency of CAT in the range of 100 mm to 1 km in length scales has already increased by almost twice \cite{storer2017}. At low altitudes, perturbations take the form of crosswinds and a variety of turbulence scales, particularly in urban environments, with structures ranging from milimeters to the scale of the largest component of the urban topology. Since these large structures are of the order, if not larger, than the length scales of the vehicles, they can destabilise the whole vehicle, leading to catastrophic events. These perturbations will also impact the robustness of ground vehicles and renewable energy systems.

\begin{figure}[t]
    \centering
    \includegraphics[width = 0.99\textwidth]{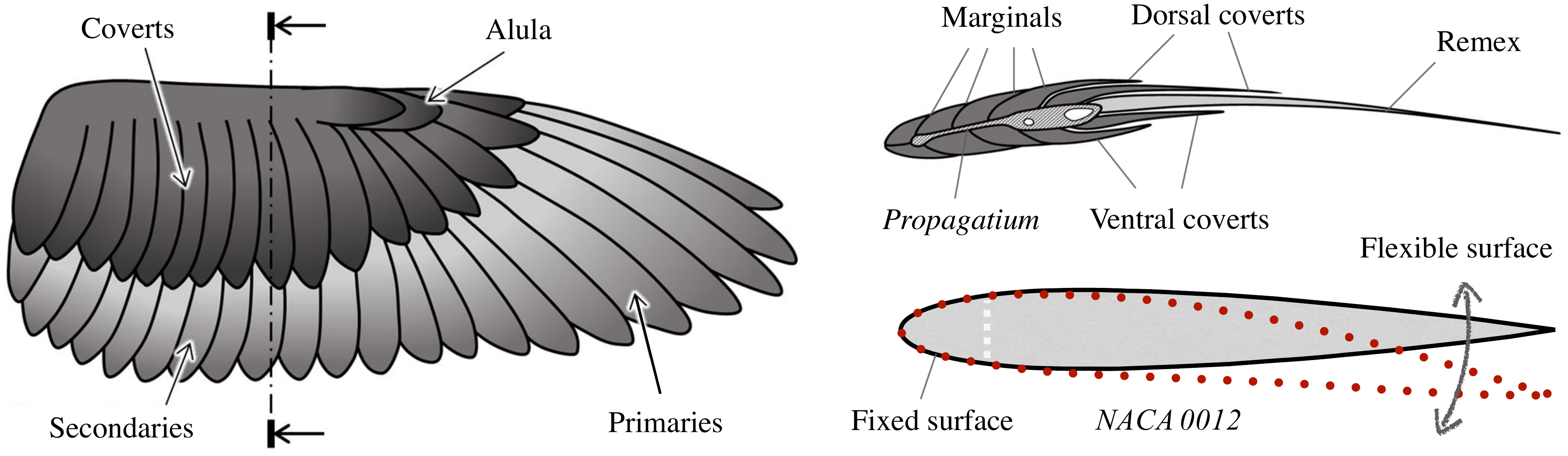}
    \caption{Schematic diagram of (a) the top view and (b) the cross-section of a bird wing. Figure adapted from \cite{Murayama}. (c) Bio-inspired flexible model for surface flexibility.}
    \label{fig: intro}
\end{figure}

Whilst man-made aerial vehicles face strict limitations during these harsh atmospheric conditions, birds can handle them and, in many cases, exploit these perturbations to cover long distances through dynamic soaring, as seen in albatrosses \cite{sachs2013}, or to gain altitude through thermal soaring, as observed in raptors \cite{Gowree2018}. Furthermore, birds can tolerate and alleviate the undesirable aerodynamic loading imposed by the unsteady flow field through the complex musculoskeletal structure of their wings~\cite{shyy_lian_tang_viieru_liu_2007, abbasi2019bio}. For example, recent studies have shown that barn owls can handle vertical gusts by changing the pitch and dihedral angle of wings through deflection at the shoulder \cite{Cheney}. 
Birds can also control flow perturbations by exploiting the transmissibility and flexibility of their feathers, which can serve as both sensors and control mechanisms. Scans of the wings sections, as illustrated in Figure \ref{fig: intro}, show that the aft $50\%$ of the wings section is very thin due to the alignment of the feathers. In this region, feathers possess varying degrees of flexibility, which potentially enhances overall lift characteristics by flapping or popping up \cite{Schluter}. The vibration or flutter of feathers under aerodynamic loading has also been demonstrated and characterized for hummingbirds in wind tunnel experiments \cite{Clark}. This characteristic can also contribute to reducing the oscillation of the wing structure by locally altering the flow over the wing or by disrupting the local vortex shedding responsible for the unsteady aerodynamic loading \cite{Murayama}. 

The aforementioned observations collectively suggest that the flexibility of the feathers acts to control the unsteady flow. This could potentially promote boundary layer transition to delay separation or help to control the shed vorticity expected at low and moderate Reynolds numbers. However, at this stage, questions still remain as to whether this type of control in birds is achieved passively, actively, or a combination of both. Drawing inspiration from these observations, applications in drone models have demonstrated that highly flexible wings can help alleviate short-period oscillations generated by gusts \cite{Murayama, zhang2018} or delay flow separation \cite{Kan2020}. On the other hand, the integration of active control mechanisms is also a promising avenue for realising these morphing capabilities. For instance, the use of Shape Memory Alloys (SMA) \cite{bil2013,barbarino2014, costanza2020} and piezoelectric actuators \cite{vos2007, henry2019} can offer precise, real-time adjustments to wing shape and surface properties, which has the potential to enhance aerodynamic forces and flight efficiency. However, their application is not without challenges, as the use of these mechanisms increases the manufacturing cost and constrains the operating temperature, among other drawbacks \cite{sofla2010}.




Furthermore, to implement appropriate flow control techniques, it is essential to first understand the behaviour of the flow when subjected to the type of vibrations expected in nature. This encouraged us to investigate how a flexible surface alters the flow as a passive control mechanism when exposed to a streamwise gust or a surging-flow type of perturbation. The investigation was divided into two parts; first, the unsteady flow on a rigid wing was studied experimentally and numerically, and the second part dealt with the impact of introducing flexibility into the wing, as introduced in Figure \ref{fig: intro}. Due to the complexity in developing a flexible-wing wind-tunnel model, this part was confined to a numerical approach.

\section{Methods}

In this section, we present the experimental framework used to validate the numerical simulations for the rigid wing, along with the methodology for incorporating flexibility into the wing numerically. For the rigid-wing case, we also discuss the principal potential-flow theories applicable to the analysis of surging flows. 

\subsection{ Experimental set-up and gust characteristics} \label{sec: Exp setup}

\begin{figure}[h!]
    \centering
    \begin{tikzpicture}
        \node[anchor=south west,inner sep=0] (image) at (0,0) {\includegraphics[trim={0 0 1.7cm 0}, clip, width=0.9\textwidth]{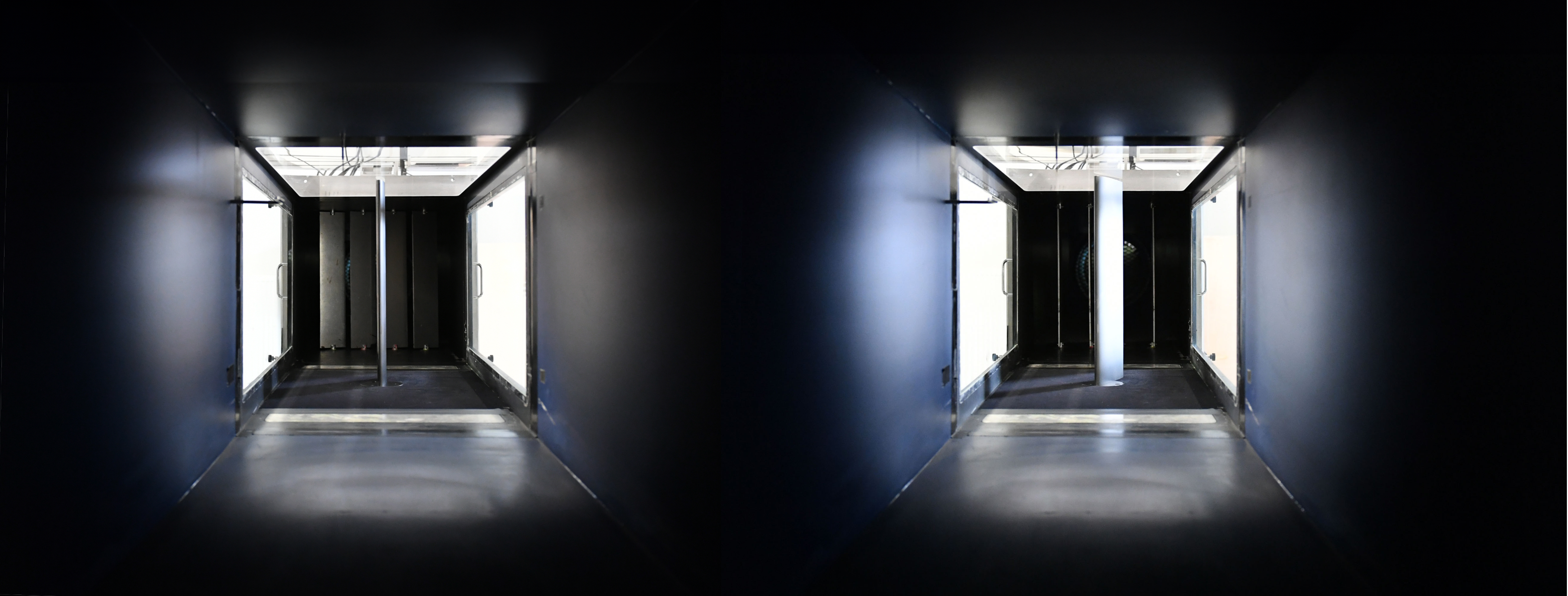}};
        \begin{scope}[x={(image.south east)},y={(image.north west)}]
            \node[anchor=north west,inner sep=7pt,text=white] at (0,1) {};
            \node[anchor=north west,inner sep=7pt,text=white] at (0.475,1) {};
        \end{scope}
    \end{tikzpicture}
    \caption{Experimental set-up showing the wing model at two different incidences and the array of louvre-doors placed downstream in (left) closed and (right) open conditions.}
    \label{fig:wind tunnel}
\end{figure}

The experimental results of Ferrand and Gowree \cite{ferrand_gowree_2022} were used as reference for the validation of the numerical simulations. The experimental campaign was conducted in the ISAE-SUPAERO open circuit wind tunnel, which features a square test section of $0.45\,\rm{m}$ and a length of $3\,\rm{m}$. The tunnel has an approximate turbulence intensity of $0.3\%$. Unsteady aerodynamic force measurements were conducted on a rigid rectangular NACA 0012 wing extending toward the extremities of the wall to approach 2D conditions. In this campaign, Louvre doors placed downstream of the test section were operated at a frequency $F_g=1.3\,\rm{Hz}$ to create an oscillating free stream by generating a varying total pressure loss downstream. Figure \ref{fig:wind tunnel} illustrates the experimental setup with the wing model at two different incidences and the Louvre doors in closed and open conditions.

\begin{figure}[t!]
    \centering
    \includegraphics[width = \textwidth]{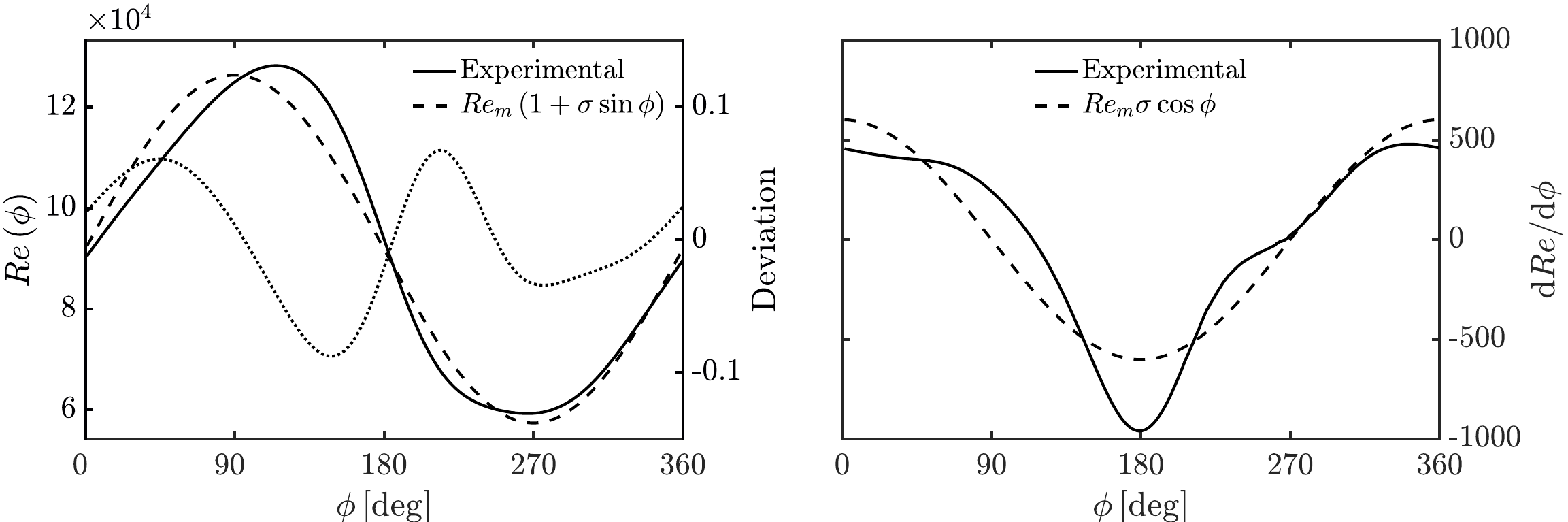}
    \caption{Unsteady free stream (left) Reynolds number and (right) acceleration profiles during the periodic gust cycle. The continuous line represents the experimental gust, while the dashed line corresponds to the theoretical sinusoidal evolution of equivalent amplitude $\sigma$. The relative deviation between both $1-u(\phi)/u_{\rm{th}}(\phi)$ is shown in the right axis of the left panel with a dotted line.}
    \label{fig:velocity acceleration}
\end{figure}

The oscillating free stream was designed to represent an incoming flow perturbation or gust in the streamwise direction, where the magnitude of the velocity varied as a sinusoidal wave with frequency $F_g$. The mean flow speed during testing was $u_m = 9.57\,\rm{m/s}$, leading to a transitional flow regime with a chord-based Reynolds number, $Re$, varying from $Re_{min} = 60\,000$ to $Re_{max} = 140\,000$. The reader is referred to Ref. \cite{ferrand_gowree_2022} for a detailed discussion of the experimental setup and measurements.
Figure~\ref{fig:velocity acceleration} depicts the velocity and acceleration profiles of the gust cycle generated in the experiments and the theoretical equivalent. The theoretical oscillating free stream profile is represented by
\begin{equation} 
u(\phi) = u_m (1+\sigma\sin{\phi}),
\end{equation}
where $u(\phi)$ represents the velocity of the oscillatory free-stream characterised by a steady velocity component $u_m$ and an oscillatory component scaled with the velocity amplitude, $\sigma$. In this case, $\phi$ denotes the different phases and can be defined as a function of the frequency of oscillations, $\omega = 2 \pi F_g$, as $\phi = \omega t$. The acceleration profile is then defined as 
\begin{equation}
    \frac{{\rm{d}}u(\phi)}{{{\rm{d}}\phi}} = u_m \sigma \cos{\phi}.
\end{equation}
The reduced frequency $k = \omega c / 2 u_m = 0.064$ and the velocity amplitude $\sigma = 0.37$ were kept the same for both the experimental and theoretical profiles. The comparison between the experimental profiles and a pure sine wave shows that there is a maximum deviation in $1-u(\phi)/u_{\rm{th}}(\phi)$ at two locations near $180^\circ$. This deviation stems from the asymmetric nature of the experimental gust profile, a result of the hysteresis tied to the flow-separation phenomenon observed on Louvre doors (flat plates with a blunt leading edge) during the high angle of incidence experienced throughout the rotational cycle.

Furthermore, based on the acceleration profile in Figure \ref{fig:velocity acceleration}, we can note two different conditions: a favourable pressure gradient due to acceleration during $\phi \in [0^\circ, 90^\circ] \cup [270^\circ, 360^\circ]$ and an unfavourable pressure gradient due to deceleration during $\phi \in [90^\circ, 270^\circ]$. The latter has the largest effect at $\phi = 180^\circ$. These conditions interact together with the Reynolds number effects ($Re < Re_m$ for $\phi \in [0^\circ, 180^\circ]$ and $Re > Re_m$ for $\phi \in [180^\circ, 360^\circ]$). Therefore, four important regions can be identified: (\textit{i}) $Re > Re_m$ with favourable pressure gradient for $\phi \in [0^\circ, 90^\circ]$, (\textit{ii}) $Re > Re_m$ with unfavourable pressure gradient for $\phi \in [90^\circ, 180^\circ]$, (\textit{iii}) $Re < Re_m$ with unfavourable pressure gradient for $\phi \in [180^\circ, 270^\circ]$, and (\textit{iv}) $Re < Re_m$ with favourable pressure gradient for $\phi \in [270^\circ, 360^\circ]$.

\subsection{Numerical set-up}

\subsubsection{Rigid wing}

The computational domain used in the numerical simulations for the rigid case consists of a two-dimensional NACA 0012 aerofoil. The domain extends $20c$ in every direction, where $c$ represents the aerofoil chord. The Reynolds number oscillates in the transition regime between $53\,000<Re<140\,000$, in accordance with the experimentally measured speed profile in Figure \ref{fig:velocity acceleration}. \texttt{ICEM CFD} software was used to generate a structured mesh over the aerofoil. A block-structured mesh arrangement with quadrilateral elements was adapted to the computational domain. Particular attention was paid to the near-wall region to be able to accurately resolve the boundary layer and the viscous sub-layer, which is of extreme importance to properly model the laminar-to-turbulent transition at this Reynolds number regime. The thickness of the wall cells was calculated to target a $y^+$ value {below 1}, and a mesh convergence study showed that a growth rate of 1.05 was necessary from the wall cell in the direction normal to the wall. A fine mesh was applied for the far-wake region in order to resolve the effect of the vortex shedding, and the remaining region of the domain was constructed with a coarser mesh. A total of $91\,000$ cells were used for the mesh, with 225 cells along the aerofoil chord.

The incompressible, unsteady Reynolds-averaged Navier--Stokes (URANS) equations were solved using the commercial software \texttt{STAR-CCM+}. An improved version of the standard $k-\omega$ model, the SST turbulence model \cite{kwSST1994}, was used in conjunction with the transition model $\gamma - {Re}_{\theta_t}$ \cite{Langtry2009} to account for the effects of boundary layer transition  at these Reynolds numbers. An implicit scheme was used with a time step of $\delta_t = 10^{-4}$ seconds, defined as a combination of the convective time, the Courant number (CFL), and the vortex shedding frequency. A total of $100$ points were used to discretise the convective time on the aerofoil, aiming to keep the CFL values below 30. From the experimental results, the vortex shedding frequency was known; therefore, $\delta_t$ was selected to reliably represent these oscillations, \textit{i.e.} at least $50$ points were resolved for each oscillation. Finally, the segregated flow solver with constant density was used to account for the incompressible nature of the flow. Mesh and time independence studies were performed to ensure the reliability of the results. {The selected mesh and time step provided a good agreement with experimental findings while correctly predicting the high-frequency effects of the boundary layer dynamics.}

\subsubsection{Flexible wing}

\begin{figure}[t!]
    \centering
    \includegraphics[width = \textwidth]{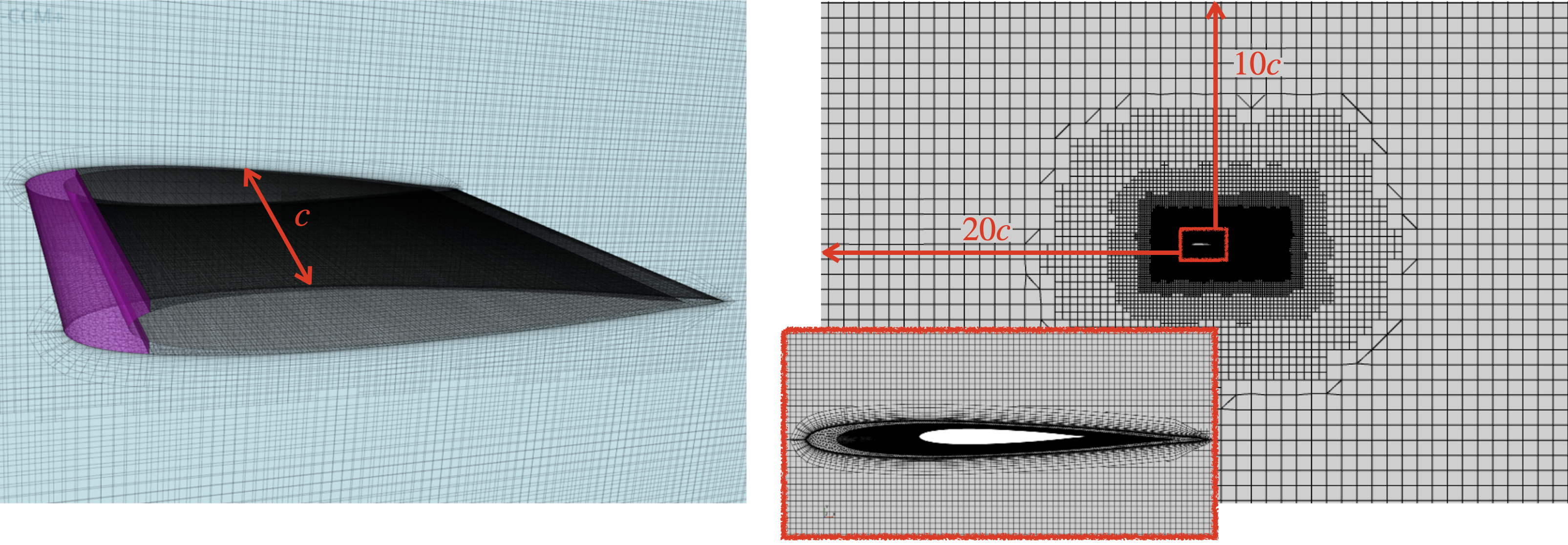}
    \caption{Left: Full simulation domain with the different meshes: the trimmed fluid mesh in the far field, the prismatic layers that discretise the boundary layer and the structural mesh. Right: Full simulation domain and a zoom in the wing region.}
     \label{fig: FSI_mesh}
\end{figure}

For the flexible case, the scheme utilised to resolve the fluid domain remains consistent with the approach used in the rigid-wing case. However, to accommodate the dynamics of the fluid-structure interactions (FSI), the domain is extended in the spanwise direction by a measure equal to one chord length. To ensure that the simulation mimics the characteristics of a two-dimensional analysis, {periodic} boundary conditions are imposed on the sidewalls. Furthermore, the FSI case also required a mesh for the structure of the wing in order to compute the loads the flow exerts on the wing and vice versa. Here, a portion of the leading edge of the wing remains fixed, whereas the rest of the structure is free to deform under the effect of aerodynamic loads. These regions are represented in Figure \ref{fig: FSI_mesh}, where the purple section corresponds to the fixed structure and the black section is free to move and deflect. The thickness of the wing skin surface in the latter section is $1\,\rm{mm}$. 

Therefore, since mesh motion is introduced, conservation equations for both fluid and solid mechanics need to be modified. For the fluid, mesh motion introduces an additional flux in the convective terms. The governing equations for the fluid and for the moving meshes can be then written as:
\begin{equation}
    \frac{\partial}{\partial t} \int_V \rho dV + \oint_A \rho \left(\boldsymbol{u}_r - \boldsymbol{u}_g\right) d\boldsymbol{a} = \int_V S_u dV,
\end{equation}
\begin{equation} \label{eq: momentum equation}
    \frac{\partial}{\partial t} \int_V \rho \boldsymbol{u} dV + \oint_A \rho \boldsymbol{u} \otimes \left(\boldsymbol{u}_r - \boldsymbol{u}_g\right) d\boldsymbol{a} = \int_A \boldsymbol{\sigma} d\boldsymbol{a} + \int_V \boldsymbol{f}_b dV - \int_V \rho \boldsymbol{\omega} \times \boldsymbol{u} dV,
\end{equation}
where $\rho(\boldsymbol{x},t)$ represents the fluid density at position $\boldsymbol{x}$ and time $t$, $V$ denotes the control volume, $\boldsymbol{a}$ is the control surface vector, $\boldsymbol{\sigma}$ the viscous stress tensor, $\boldsymbol{f}_b$ denotes the resultant of body forces and $S_u$ represents a source term. Moreover, $\boldsymbol{u}(\boldsymbol{x},t)$ is the velocity in the fixed reference frame (also called the absolute velocity), $\boldsymbol{u}_g$ is the grid velocity and $\boldsymbol{u}_r$ denotes the relative velocity. The latter velocity terms are introduced as a result of the mesh motion; if the mesh reference frame is not moving, the grid velocity vanishes and $\boldsymbol{u} = \boldsymbol{u_g}$. Additionally, the non-inertial moving reference frame introduces a new term on the right-hand side of Equation \ref{eq: momentum equation}, which represents a fictitious force consisting of Coriolis and centrifugal forces. 

Since the mesh is moving, the shape and position of its cells can vary with time. In such a case, an additional equation needs to be solved to enforce space conversation:
\begin{equation}
    \frac{\partial}{\partial t} \int_V dV = \int_A \boldsymbol{u}_g d\boldsymbol{a}.
\end{equation}
This ensures that the rate of change of a cell volume balances the motion of its bounding surface. The two-way coupling solver was selected for data exchange between the fluid and solid solvers: in addition to the interaction of the fluid on the solid, the interaction of the solid displacement on the flow is also considered by morphing at each time step the fluid mesh according to the fluid/solid interface deformation. The discretisation scheme is a finite-volume method (FVM) for the fluid part and finite-element method (FEM) for the solid part. Figure \ref{fig: FSI_mesh} also illustrates the mesh used in the fluid-structure-interaction case, comprising $1.37$ million cells for the fluid and $412\,000$ cells for the wing structure. The simulations were configured for parallel processing using the high-performance computing resources at ISAE-SUPAERO. Each of the cases was executed using a total of 192 cores operating at 2.6 Hz, which achieved an approximate computational performance of 0.99 teraflops.

\subsection{Unsteady lift predictions from potential flow theory}


When an aerofoil is exposed to an oscillating free stream velocity $u(t)$, it induces a variation in the bound aerofoil circulation $\Gamma(t)$, generating unsteady lift even when the angle of attack $\alpha$ is constant. The unsteady lift overshoot is classically examined using potential flow theories by Isaacs~\cite{isaacs1945} and Greenberg~\cite{greenberg1947}, which assume a large fore-aft motion wavelength compared to the chord length $c$ and a small angle approximation $\sin(\alpha) \approx \alpha$. The velocity amplitude is assumed to be $\sigma<1$ to prevent reverse flow. The unsteady lift is normalised with the steady lift, $L_s = \rho \pi u_m^2 \alpha$, of a flat plate in potential flow, where $\rho$ is the air density.

This study uses the ratio of non-dimensional lift coefficients for comparison with experimental and numerical results, following Strangfled \textit{et al.} \cite{strangfeld2016}. All dynamic effects are quantified using the non-dimensional lift coefficient ratio as shown in Equation \ref{eq: cl_clqs}, proposed by Isaacs~\cite{isaacs1945} and suggested by Ref.~\cite{van1994}:
\begin{equation}
\label{eq: cl_clqs}
\frac{C_l(t)}{C_{l,qs}} = \frac{L(t)}{L_s}\frac{1}{(1 + \sigma \sin(\omega t))^2}.
\end{equation}
Isaacs theory, which makes no additional assumptions about the wake, is considered mathematically exact \cite{van1992}. It is represented in Equation \ref{eq: Isaacs}, where $0.5 k\cos(\omega t)$ is the non-circulatory solution, and all other terms represent the circulatory solution, including the quasi-steady lift and the unsteady wake denoted by $l_m$. The coefficients $l_m$ are defined as a combination of the real and imaginary parts of the Theodorsen function $C(nk)$:
\begin{equation}
\label{eq: Isaacs}
\begin{aligned}
& \frac{C_l(t)}{C_{l, q s}}= \frac{1}{(1+\sigma \sin (\omega t))^2}\left[1+0.5 \sigma^2+\sigma\left(1+\operatorname{Im}\left(l_1\right)+0.5 \sigma^2\right) \sin (\omega t)\right.  \\ 
& \left.+\sigma\left(\operatorname{Re}\left(l_1\right)+0.5 k\right) \cos (\omega t)+\sigma \sum_{m=2}^{\infty}\left(\operatorname{Re}\left(l_m\right) \cos (m \omega t)+\operatorname{Im}\left(l_m\right) \sin (m \omega t)\right)\right].
\end{aligned}
\end{equation}
Conversely, Greenberg's theory, which models the wake vorticity in a sinusoidal harmonic form, leads to the normalised lift coefficient in Equation \ref{eq: Greenberg}. Here, $F$ and $G$ are defined as a combination of the real and imaginary parts of the Theodorsen function $C(nk)$:
\begin{equation}
    \label{eq: Greenberg}
    \begin{aligned}
    \frac{C_l(t)}{C_{l, q s}}= & \frac{1}{(1+\sigma \sin (\omega t))^2}\left[\left(1+0.5 \sigma^2 F\right)+\sigma(1+F) \sin (\omega t)\right. \\
    & \left.+\sigma(0.5 k+G) \cos (\omega t)+0.5 \sigma^2 G \sin (2 \omega t)-0.5 \sigma^2 F \cos (2 \omega t)\right].
    \end{aligned}
\end{equation}
Despite its simplicity, Greenberg's high-frequency assumption limits its application range, as it neglects the wake flow oscillation
, leading to an underestimation of the reduced frequency for lower velocities and an overestimation for higher velocities. In the present work, the accuracy of both theories to predict the aerodynamic behaviour of aerofoils under the effect of surging flows with non-negligible viscous effects will be assessed. 

\section{Rigid wing}

\subsection{Steady free stream} \label{sec: rigid steady}

\begin{figure}[t!]
     \centering
     \includegraphics[width=0.48\textwidth]{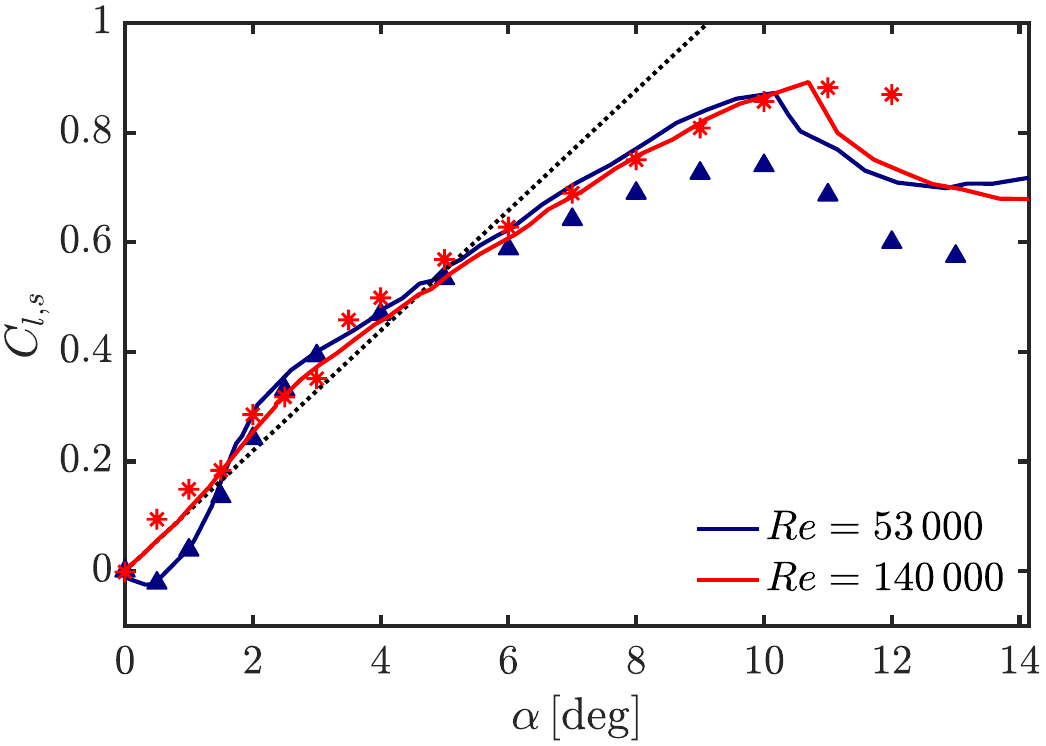}
     \hfill
     \includegraphics[width=0.49\textwidth]{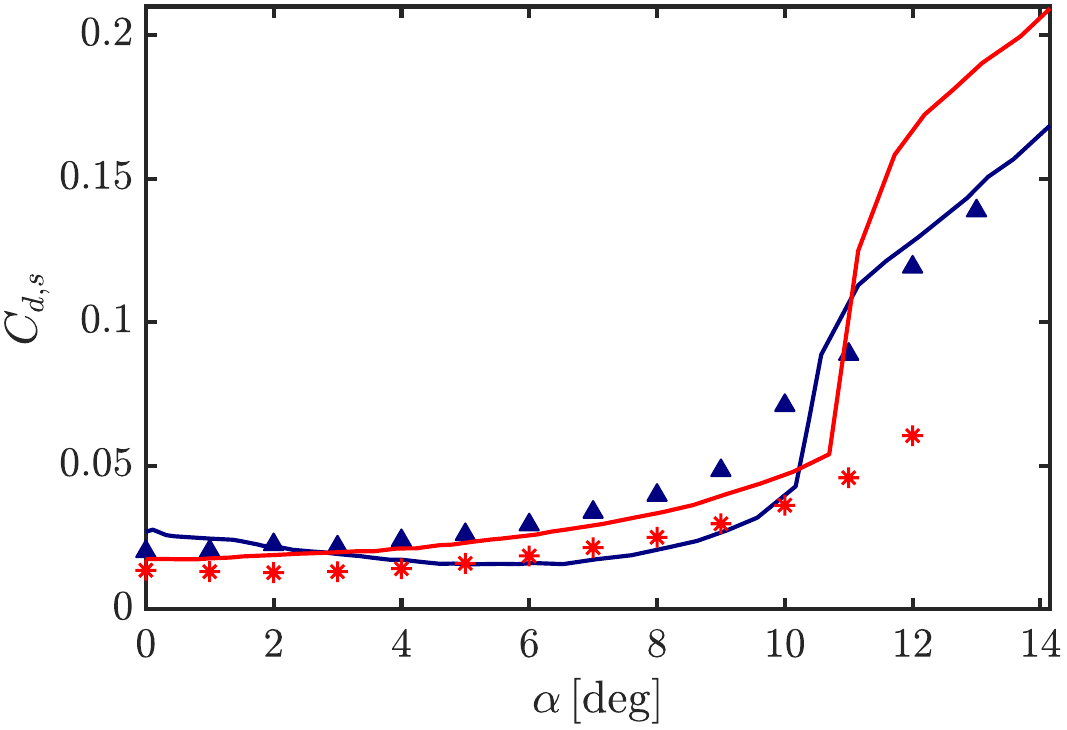}
     \caption{Static lift and drag coefficients with angle of attack $\alpha$. Lines denote experimental data from ISAE-SUPAERO open-circuit wind tunnel whereas markers show numerical results for two different Reynolds numbers. The thin aerofoil theory prediction $C_l = 2 \pi \alpha$ is reported on the left plot.}
     \label{fig: Curves NACA0012}
\end{figure}

Figure \ref{fig: Curves NACA0012} illustrates the comparison between experimental and numerical results for the lift and drag coefficients under steady free-stream conditions in two different Reynolds numbers, $Re = 53\,000$ and $Re = 140\,000$. These Reynolds numbers were chosen to represent the minimum and maximum velocities encountered in the streamwise gust. Under these conditions, the aerodynamic performance of the aerofoil is significantly influenced by laminar separation or the presence of a recirculating laminar separation bubble (LSB). Moreover, the size of the separated region delineates two distinct flow regimes: short- and long-bubble regimes \cite{Tani1964}. The existence of these regimes alters the effective shape of the aerofoil, consequently impacting its aerodynamic characteristics. Specifically, the extended separation region affects the pressure distribution, leading to nonlinearities in the overall aerodynamic performance, as depicted in Figure \ref{fig: Curves NACA0012} within the pre-stall $C_{l,s}$ region. At $Re = 53\,000$, the flow exhibits a higher viscous nature, resulting in a greater deviation from the predictions of the thin-aerofoil inviscid theory for angles of attack $0^\circ<\alpha <3^\circ$. This $Re$ is more prone to long-bubble regimes without reattachment, in contrast to $Re=140\,000$, which is associated with short-bubble regimes and subsequent flow reattachment. {Despite these complex phenomena, a notable agreement between numerical and experimental data is evident in the pre-stall region, which is even more pronounced for \(Re=140\,000\), underscoring the efficacy of the \(\gamma-Re_{\theta_t}\) transition model for these flow conditions.}

\begin{figure}[t!]
     \centering
     \includegraphics[width=\textwidth]{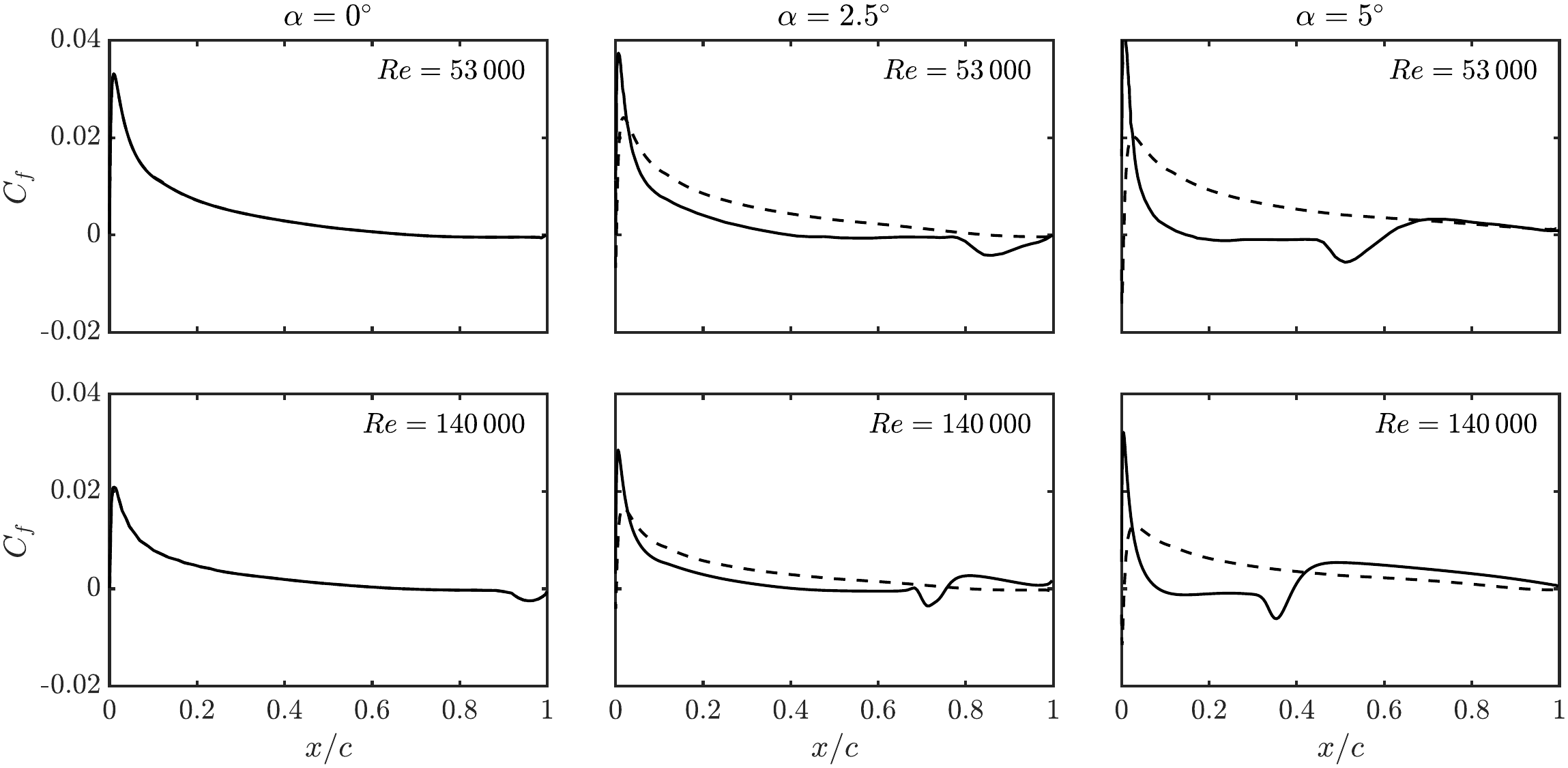}
     \caption{Chordwise evolution of time-averaged skin-friction coefficient $C_f$ distributions at different incidence angles for $Re = 53000$ and $Re = 140000$. The upper surface of the aerofoil is illustrated in solid line, while the lower surface is shown in dashed line.}
     \label{fig: Cf_Re_Static}
\end{figure}

The impact of the laminar separation bubble on the aerodynamic behavior of the airfoil can be further elucidated by examining the evolution of the skin friction coefficient $C_f$. Figure \ref{fig: Cf_Re_Static} compares the time-averaged $C_f$ across various angles of attack $\alpha$ and Reynolds numbers $Re$. The LSB is identifiable as the region where the skin friction coefficient remains negative. Initially, the boundary layer is laminar and attached to the surface, hence $C_f > 0$. The first transition of $C_f$ from positive to negative marks the onset of laminar separation, and the subsequent flow regime depends on whether the flow reattaches as a turbulent boundary layer (long-bubble regime) or not (short-bubble regime). For the cases in Figure \ref{fig: Cf_Re_Static}, this is primarily dictated by the effect of $\alpha$ and $Re$:
\begin{itemize}
    \item At $\alpha = 0^\circ$, laminar separation without reattachment occurs on both the upper and lower surfaces of the aerofoil, independent of the Reynolds number.
    \item At $\alpha = 2.5^\circ$, the dynamics of the flow are more complex. The lower surface experiences laminar separation without reattachment for both Reynolds numbers, whereas the behaviour on the upper surface is Reynolds-number dependent. For the lower $Re$, the flow does not reattach, compared to the higher $Re$ where turbulent reattachment occurs.
    \item Beyond $\alpha = 5^\circ$, no separation is observed on the lower surface, while the upper surface transitions to a short-bubble regime for both $Re$.
\end{itemize}

This delineation of flow regimes sets the stage for a detailed investigation in subsequent sections. The case in $\alpha = 2.5^\circ$ is particularly intriguing, as it encapsulates the characteristics of both regimes. Therefore, it acts as an important test bed for assessing the accuracy of our numerical methodology and the relevance of potential flow theories when considering unsteady effects.

\subsection{Unsteady free stream}

\begin{figure}[t!]
    \centering
    \begin{minipage}{0.9\textwidth}
    \begin{minipage}{\textwidth}
    \centering
    \includegraphics[trim={0 11.5cm 0 0}, clip, width = 0.55\textwidth]{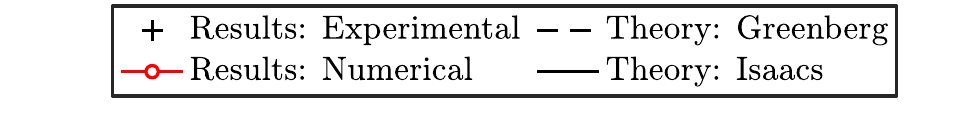}
    \end{minipage}
    \includegraphics[trim={0 0.1cm 0.5cm 0.6cm}, clip, width = 0.49 \textwidth]{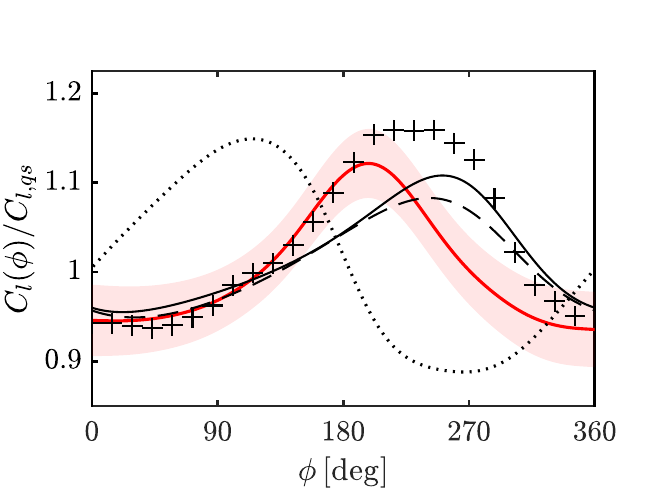}
    \hspace{0.002\textwidth}
    \includegraphics[trim={0 0.1cm 0.5cm 0.6cm}, clip, width = 0.4875 \textwidth]{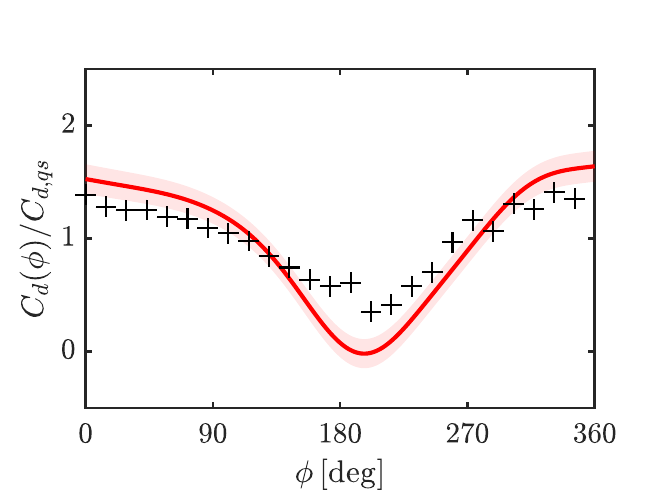}
    \end{minipage}
    \caption{Phase-averaged (left) lift coefficient ratio $C_l/C_{l,qs}$ and (right) drag coefficient ratio $C_d/C_{d,qs}$ for $\alpha = 2.5^\circ$. The evolution of the unsteady free stream along the gust cycle is represented as a dotted line on the left panel.}
    \label{fig: Cl Cd Dynamic}
\end{figure}

The goal now is to evaluate the impact of the unsteady free stream. Figure \ref{fig: Cl Cd Dynamic} represents the evolution of lift and drag coefficients during a gust cycle, which have been normalized with the corresponding values at static conditions. The angle of incidence was selected based on the flow regimes observed during the static configuration and corresponds to $\alpha = 2.5^\circ$. At this angle of incidence, the upper surface experienced laminar separation with or without reattachment depending on $Re$, while the lower surface underwent separation without reattachment. Due to the time-dependant nature of the incoming free stream, the numerical results shown in Figure \ref{fig: Cl Cd Dynamic} are acquired from a phase-averaging technique over 120 gust cycles. Although more gust cycles were assessed, no significant variations were detected. Consequently, this threshold was deemed appropriate and adopted for subsequent analyses.

We observe two different levels of agreement between the numerical, theoretical, and experimental results. During the first half of the gust, $\phi \in [0^\circ, 180^\circ]$, the numerical results show remarkable agreement with the experimental results. On the other hand, potential-theory predictions show this agreement for $\phi \in [0^\circ, 90^\circ]$ and $\phi \in [270^\circ, 360^\circ]$. This implies that the most critical region of the gust to model is $\phi \in [180^\circ, 270^\circ]$, which is dominated by an unfavorable pressure gradient due to the deceleration of the incoming free stream and larger viscous effects (i.e. $Re < Re_m$). During the wind-tunnel experiment, even if the intensity of the background perturbations is low, they will still accelerate the growth rate of disturbances within the boundary layer, leading to transition and turbulent reattachment. This effect can be resolved properly using direct numerical simulations (where all the spatial and temporal scales of the flow are resolved without any turbulence model) and linear and nonlinear stability theory. However, we hypothesize that the models used in this study do not model the amplification of these unstable modes, which plays an important role during the unfavorable pressure gradient and low-$Re$ part of the gust. Despite this limitation, our numerical simulations are able to correctly predict a change in phase of the $C_l/C_{l,qs}$ peak due to viscous effects, with respect to potential theory. The peak in the latter case occurs solely as a result of the lowest Reynolds during the gust. However, we show here that at these transitional Reynolds numbers, it is also important to consider deceleration or acceleration effects as they can promote or inhibit the growth of instabilities within the boundary layer. The latter is in line with previous experimental studies of two-dimensional airfoils under longitudinally oscillating streams \cite{strangfeld2016, Greenblatt2023}.

\begin{figure}[t!]
    \centering
    \includegraphics[trim={0 1.cm 3.25cm 2cm},clip,width = 0.4425\textwidth]{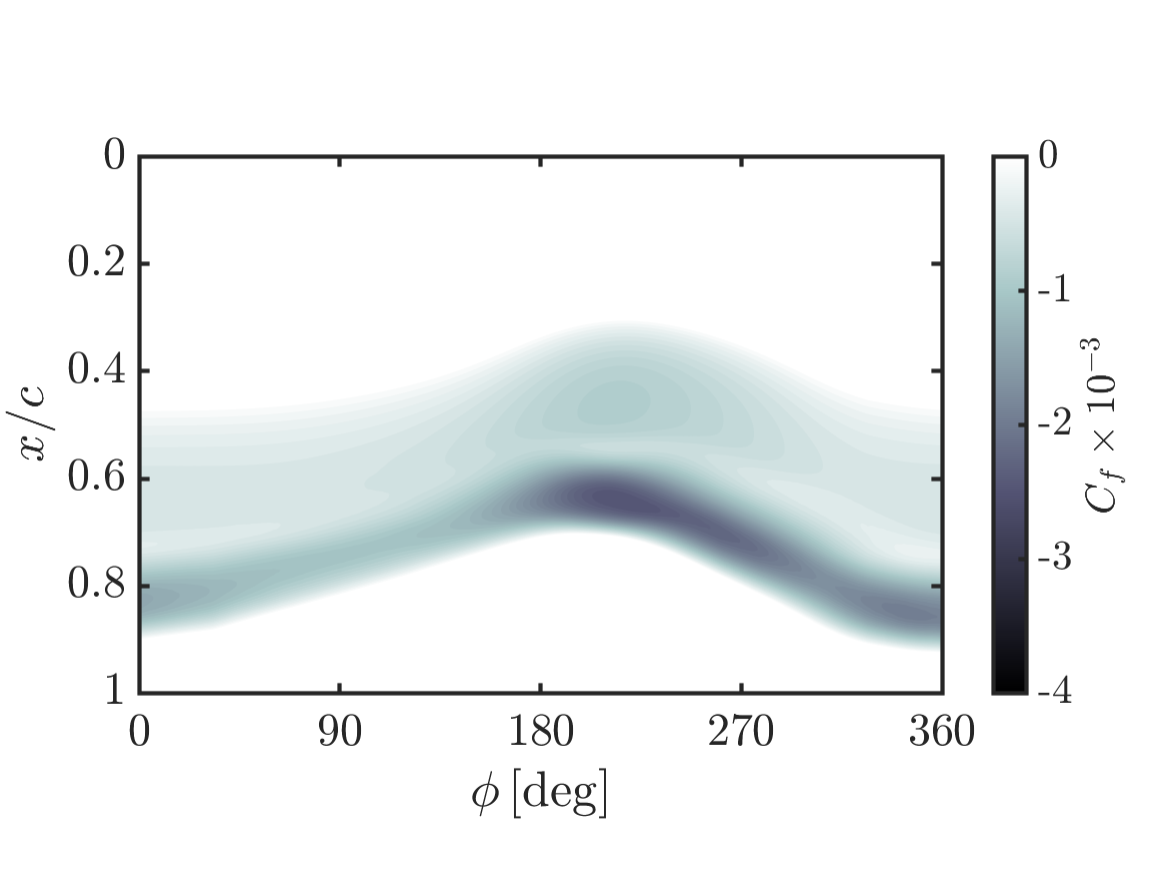}
    \hspace{0.04\textwidth}
    \includegraphics[trim={2.15cm 1.cm 0.2cm 2cm},clip,width = 0.4675\textwidth]{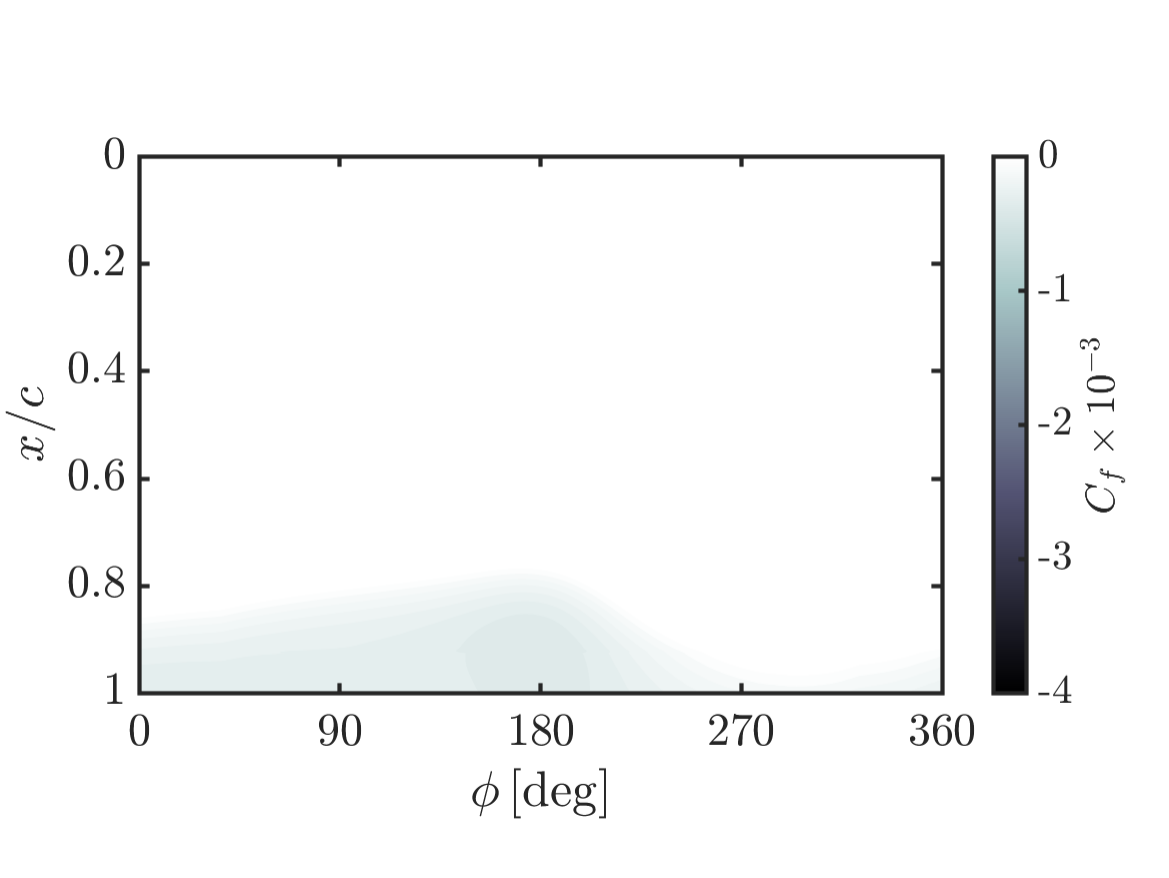}
    \caption{Phase-averaged skin-friction coefficient $C_f$ map depicted for the extrados (upper panels) and intrados (lower panels) of the aerofoil across various chordwise positions and gust locations at angles of attack $\alpha = 0^\circ$ and $2.5^\circ$. Only negative $C_f$ values are illustrated to emphasize the location of the laminar separation bubble. 
    }
    \label{fig: Cf Dynamic Map}
\end{figure}

The effect of the pressure gradient at these Reynolds numbers range can also be observed by analyzing the evolution of the phase-averaged skin-friction coefficient along the gust. Contour maps for negative values of this quantity are shown in Figure \ref{fig: Cf Dynamic Map}, which enables us to observe the evolution of the bubble regimes across different gust regions. On the upper surface, laminar separation with reattachment occurs for all $\phi$. However, for $\phi \in [180^\circ, 270^\circ]$, laminar separation occurs earlier at $x/c \approx 0.4$, and larger phase-averaged $| C_f |$ values with high-frequency oscillations are observed around $x/c \approx 0.7$. This suggests a higher instability of the boundary layer and shed vorticity effects in this gust region, which is dominated by $Re < Re_m$ and unfavorable effects of the pressure gradient. Furthermore, the oscillations found in this region of the gust are in agreement with Strangfeld \textit{et al.} \cite{strangfeld2016}, who reported a high frequency oscillation near the overshoot in lift coefficient due to the formation and shedding of a recirculation bubble near the trailing edge that modified the Kutta condition. In our study, it is the lower surface the one that exhibits some oscillations between fully-attached and laminar-separation regimes in the same gust region. The alignment of this phenomenon with the interaction between these two regimes and its true appearance in the experimental results may be the reason behind the underprediction of the experimental lift ratio curve, since the location of the separation bubble and its subsequent shedding play a decisive role in the resulting aerodynamic loads \cite{strangfeld2016}. Hence, we show that at this Reynolds number regime, the lower surface might also play a role in the prediction of aerodynamic forces.

\begin{figure}[t!]
    \centering
    \includegraphics[width = 0.55\textwidth]{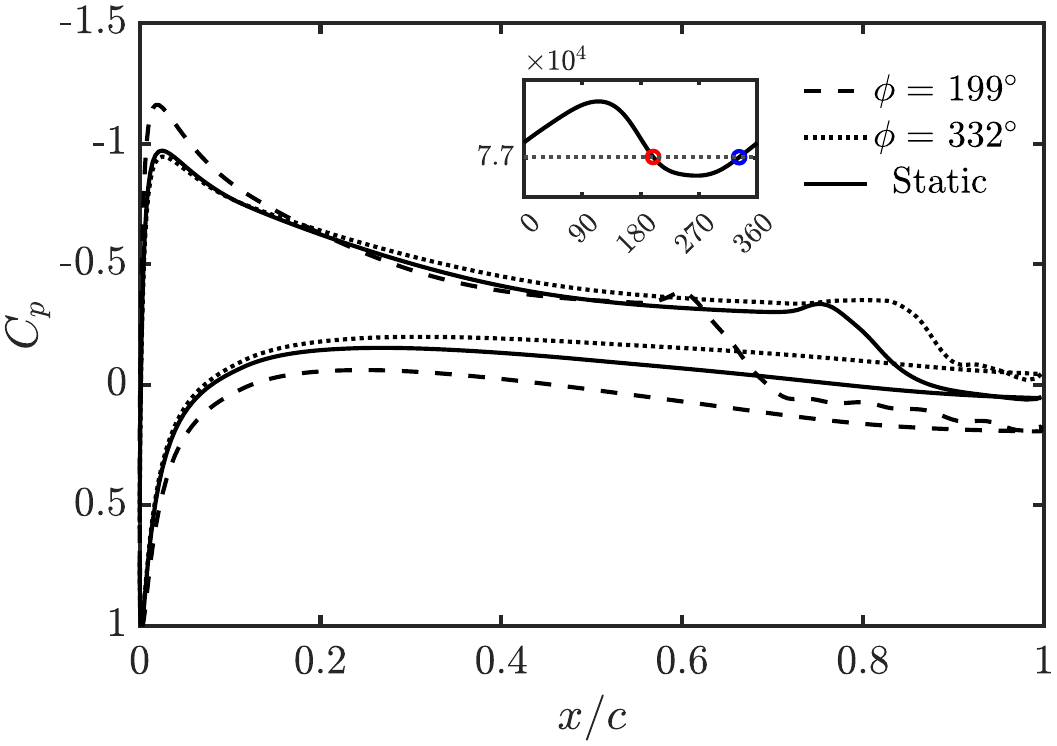}
    \caption{Chordwise evolution of pressure coefficient $C_p$ over the aerofoil for a Reynolds number $Re = 77 000$ for dynamic and static conditions at $\alpha = 2.5^\circ$. This $Re$ is achieved in the dynamic case at $\phi = 332^\circ$ and the location of maximum $C_l/C_{l,s}$, $\phi = 199^\circ$.}
    \label{fig: Cp Re 66k}
\end{figure}

To conclude the analysis, Figure \ref{fig: Cp Re 66k} shows the chordwise evolution of the pressure coefficient $C_p$ under both dynamic and static conditions at $\alpha = 2.5^\circ$ and $Re = 66,000$. During the acceleration of the flow, which occurs at $\phi = 332^\circ$ for the unsteady free stream, a favourable pressure gradient is introduced. This stabilises the boundary later by suppressing the growth of boundary layer instabilities. Conversely, the deceleration phase, occurring at $\phi = 199^\circ$, introduces an unfavourable pressure gradient, promoting flow instabilities. These instabilities, in turn, accelerate the onset of laminar separation. This is clearly observed in Figure \ref{fig: Cp Re 66k}, where for the same Reynolds number, $\phi = 199^\circ$ features an earlier separation location at $x/c = 0.6$, whereas separation occurs at $x/c = 0.85$ for $\phi = 332^\circ$.

Therefore, while the effects of Reynolds number dominate the flow behavior in the steady free stream, the findings from Figure \ref{fig: Cp Re 66k} suggest otherwise for the dynamic case. In regions where the highest $Re$ is observed, one would typically expect the earliest onset of transition and reattachment. However, due to the proximity of this region to the peak acceleration phase, the opposite is observed, with the transition and reattachment location being the most downstream. This distinction between the effects of acceleration and deceleration can be further accentuated by the observation that the length of the separation bubble is marginally elongated during the deceleration phase. This suggests that not only does the acceleration influence the onset of separation, but it also plays a role in determining the reattachment location of the flow. In the following sections, we introduce flexibility effects to the wing models to observe how the flow topology changes and whether the high-frequency oscillations can be mitigated.

\section{Flexible wing} 

\subsection{Steady free stream}

\begin{table}[t!]
\setlength{\arrayrulewidth}{0.3pt}
\centering
\resizebox{\textwidth}{!}{%
\begin{tabularx}{1.2\textwidth}{l *4{>{\centering\arraybackslash}X}@{}}
\hline
{Label} & {Material} & {Stiffness} $\left[\rm{GPa}\right]$ & {Poisson ratio} & {Density} $\left[\rm{g/m^3}\right]$  \\\hline
Aluminum  & Aluminum 6061    & 69.1  & 0.33  & 2700  \\
Composite & CYCOM\textregistered 977 & 142.7 & 0.336 & 1620  \\
Super-composite & CYCOM\textregistered 977$^\dag$ & 1427 & 0.336 & 1620  \\
\hline
\end{tabularx}
}
\caption{Material properties of the wings used in the structural analysis. The table presents the stiffness, Poisson ratio, and density for each material. $^\dag$The properties of this composite material are not realistic, representing a material that is ten times stiffer than CYCOM\textregistered 977.}
\label{tab:materials_fsi}
\end{table}

\begin{figure}[t!]
    \centering
    \includegraphics[width = \textwidth]{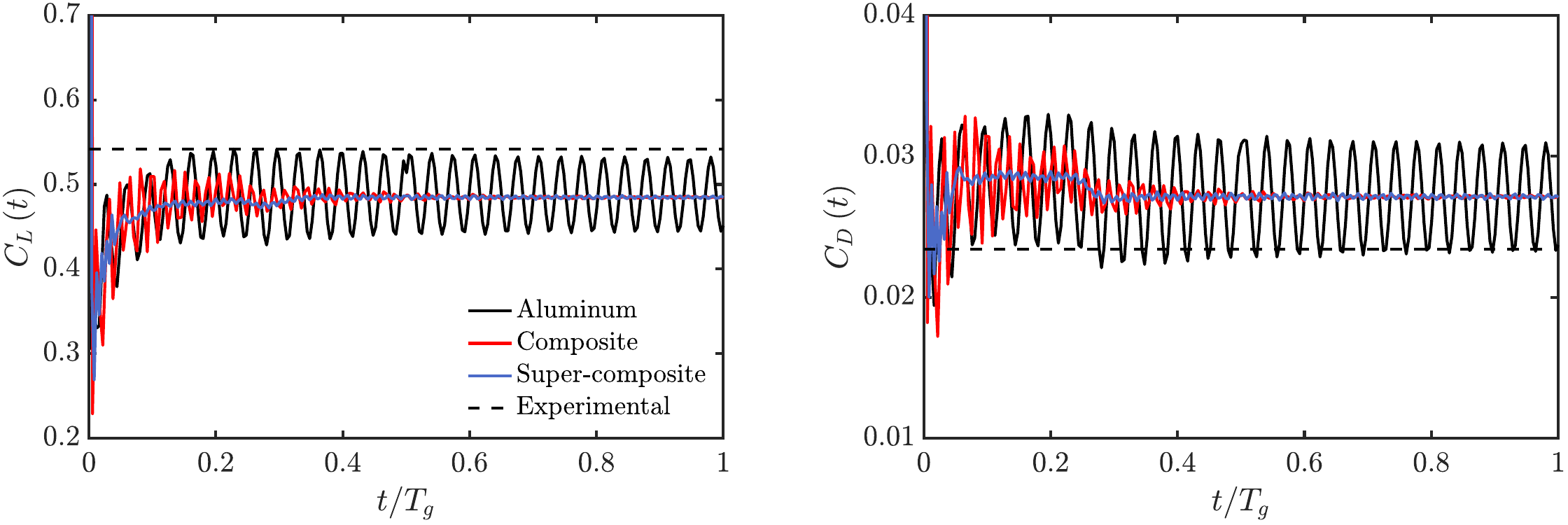}
    \caption{Temporal evolution of lift (left) and drag (right) coefficients for wings with different compositions under a steady free stream. Time is normalized using the duration of the gust cycle from the unsteady case, denoted as $T_g$.}
    \label{fig: QS_Coeffs}
\end{figure}

To start, the impact of the stready free stream on a flexible NACA 0012 wing model composed of the materials reported in Table \ref{tab:materials_fsi} was investigated at $\alpha = 5^\circ$ and $Re = 140\,000$. Figure \ref{fig: QS_Coeffs} illustrates the instantaneous $C_L$ and $C_D$ for three different wing types, categorized based on their material properties. In contrast to the pure 2D aerodynamic calculation in \S\ref{sec: rigid steady}, which closely matched the experimental results in Figure \ref{fig: Curves NACA0012}, the mean lift from the aluminum wing is slightly lower in this case. In FSI modelling, a finite 3D domain is required. Hence, in order to achieve a quasi-infinite-span condition, a periodic boundary condition is imposed. This is done to reduce the loss of lift due to downwash, which would also increase the total drag due to the generation of induced drag. Therefore, the reduction in mean lift and the increase in overall drag, in comparison to the static case, is due to the modification of the pressure distribution resulting from the modification in the topology of the laminar separation bubble and earlier re-attachment to a turbulent boundary due to the vibration of the surface, even for the aluminium case.

Interestingly, the oscillations in both $C_L$ and $C_D$ observed for the rigid wing simulated in \S\ref{sec: rigid steady} are significantly damped when flexibility is introduced into the wing structure. While the aluminum wing exhibits convergence towards an oscillatory state, the two composite wings are capable of suppressing the oscillations and approaching an almost steady behavior. These oscillations are purely due to the unsteadiness induced by the laminar separation bubble. Therefore, introducing composite materials into the structure, we observe that we can modify the topological structure of the flow to mitigate the impact of unsteadiness on aerodynamic loads. Moreover, the use of composite materials also leads to a reduced time required to reach this quasi-steady state. In fact, the super-composite wing is able to reach this state within only 20\% of the time corresponding to one gust cycle, whereas the composite wing requires a 50\%. Therefore, we note that more advanced materials with enhanced structural properties can achieve lower relaxation times. For subsequent sections, the analyses will be confined to the aluminum and composite cases only, as these scenarios employ materials with realistic physical properties. Given the behavior observed in the steady state for the super-composite case, we assert that conclusions drawn from the composite wing could be extrapolated to other materials with similar structural properties.

\subsection{Unsteady free stream}

\begin{figure}[t!]
    \centering
    \includegraphics[trim={2.5cm 2.9cm 3.1cm 3.5cm}, clip,width = \textwidth]{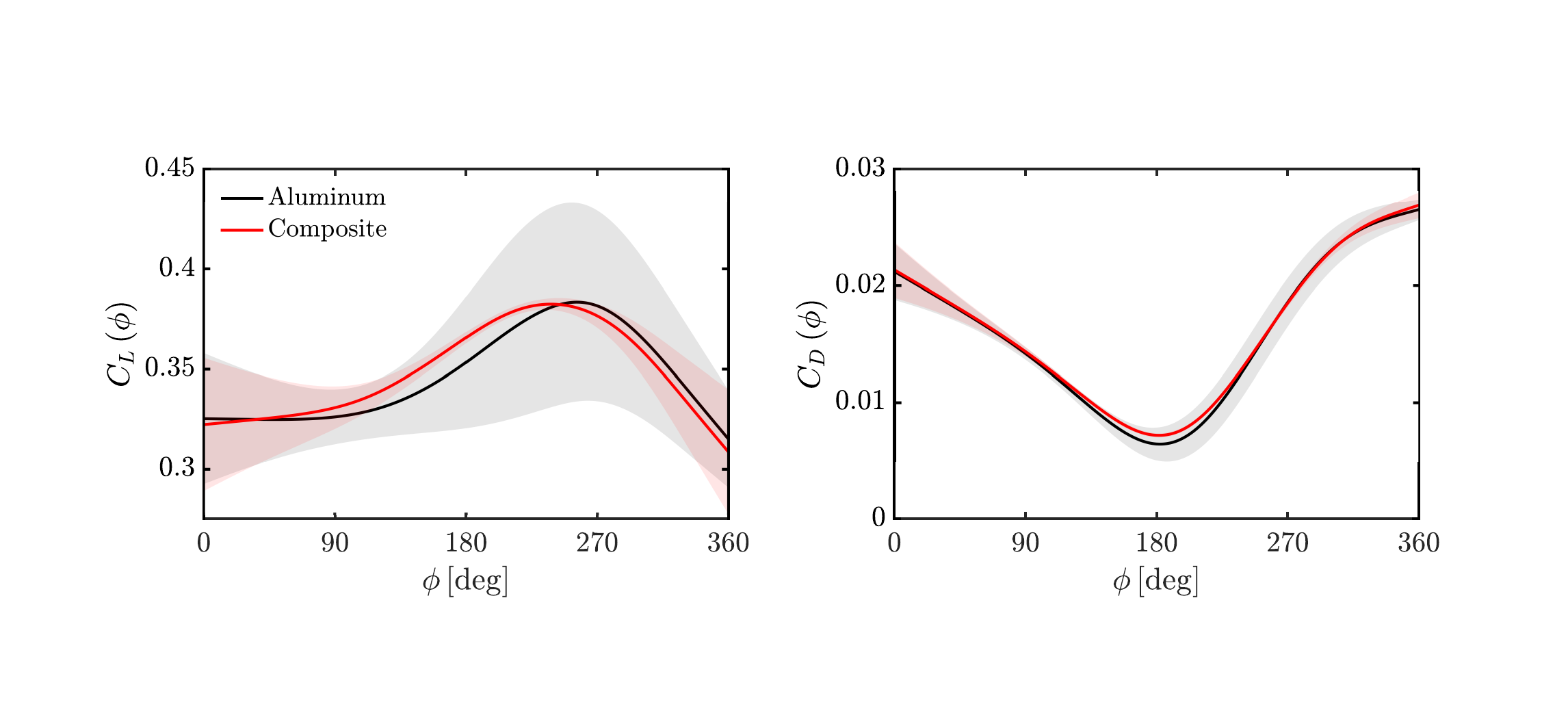}
    \caption{Lift (left) and drag (right) coefficient comparison between the phase-averaged numerical results from the (black) aluminum and (red) composite wings for $\alpha=2.5^\circ$. The oscillation amplitude during the gust cycle is shown with shaded colors.}
    \label{fig: CL_CD_uns_FSI}
\end{figure}

Figure \ref{fig: CL_CD_uns_FSI} showcases the phase-averaged unsteady $C_L(\phi)$ and $C_D(\phi)$ together with their corresponding oscillation amplitudes for both the aluminium and composite wing surfaces. 
In the initial phase of the cycle, specifically for $\phi<120^\circ$, the mean values of $C_L$ from both numerical cases exhibit only minor deviations. This behavior aligns well with the results from the 2D pure aerodynamic calculations depicted in Figure \ref{fig: Cl Cd Dynamic}. However, there is a discernible shift in the peak of the lift amplitude with respect to two-dimensional results. 
The observed deviations in the aerodynamic coefficients highlight the existence of a fluid-structure interaction mechanism, which consequently modulates the unsteady flow characteristics.
Intriguingly, this interaction is most pronounced in the region $90^\circ<\phi<270^\circ$, corresponding to the decelerating phase of the cycle. This phase is critical as it can significantly influence flow stability. Regarding the drag coefficient, the overall trend captured by the FSI simulations aligns with the results from the two-dimensional simulations.

From the analysis presented, it is evident that the flexibility of the wing plays an important role in stabilizing the flow and damping the oscillations observed in the aerodynamic forces. This stabilizing effect of flexibility is particularly pronounced during the deceleration phase of the gust. During this phase, the flow tends to promote a separation closer to the leading edge compared to what is observed during acceleration. This behavior might elucidate why, during acceleration, the oscillation magnitude for the composite case remains comparable to that of the rigid wing. In essence, the inherent flexibility of the wing serves as a mitigating factor against the adverse effects of flow instabilities, especially during phases of the gust that inherently promote flow separation and instability.

\begin{figure}[t!]
    \centering
    \includegraphics[trim={3.5cm 3.7cm 3.75cm 4.2cm}, clip,width = \textwidth]{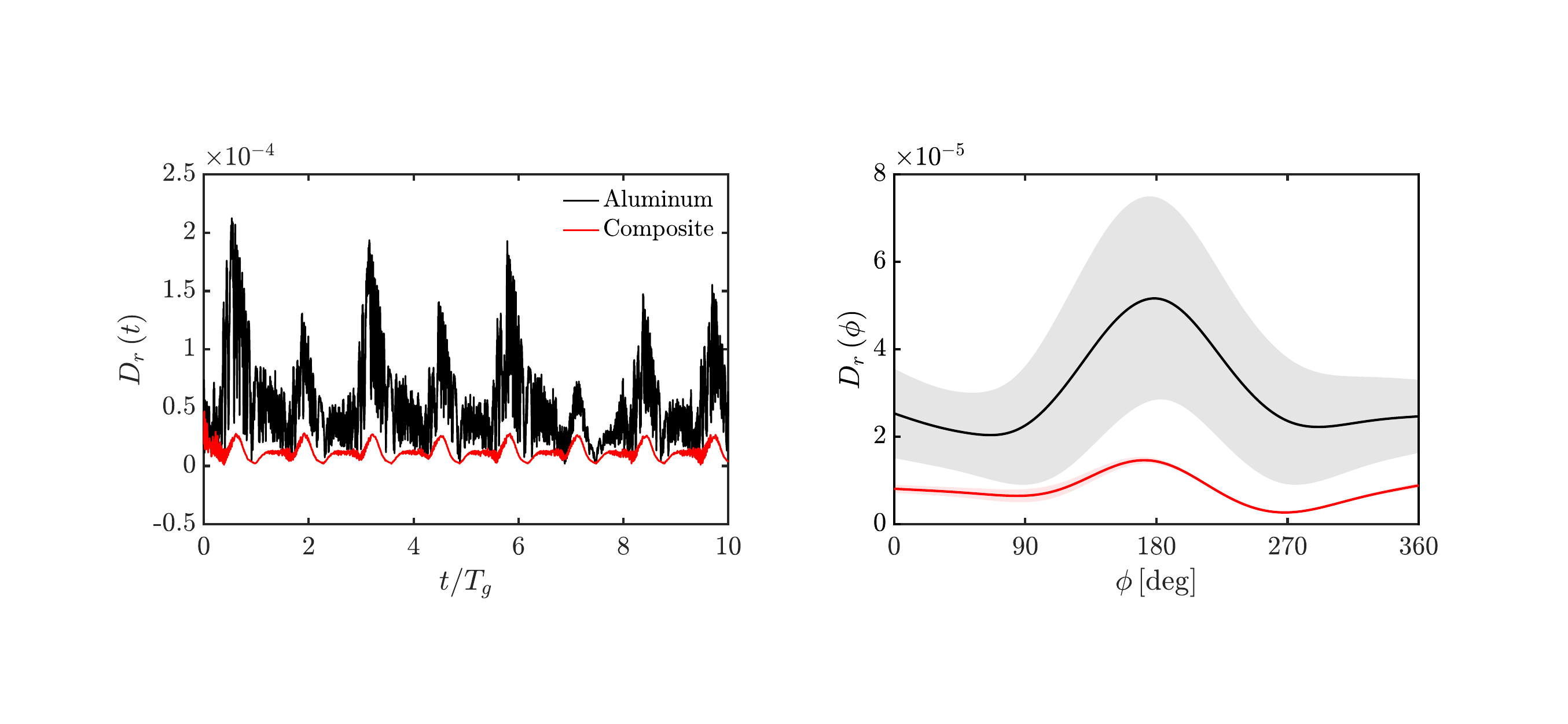}
    \caption{Comparison between the rigid wing and flexible wing of the instantaneous (left) and phase-averaged (right) relative displacements. The variable $T_g$ represents the period of the longitudinal gust.}
    \label{fig: DISP_uns_FSI}
\end{figure}

To gain further insight into the behaviour of the wing itself during the gust, Figure \ref{fig: DISP_uns_FSI} presents the instantaneous and phase-averaged values of the maximum relative displacement, $D_r$, which indicates the maximum normal distance from the non-deformed shape normalised with the chord. Remarkably, the temporal evolution of $D_r$ manifests a dissimilarity between the composite and aluminium wings. The composite wing experiences an almost uniform limit-cycle oscillation in displacement in each gust cycle after the initial transient phase, while the aluminum wing undergoes varying displacements amplitudes in each gust. The peak displacements exhibit differences of up to $80\%$, in contrast to as little as $2\%$ observed in the flexible case. In order to obtain a better understanding of the magnitude of displacement, the phase averaged values are presented on the right. This clearly depicts a substantial difference between the two materials. Despite exhibiting a similar deformation profile during the gust, the composite wing displays lower amplitude of deflection, a peculiar trait given its natural propensity to vibrate freely. Most importantly, the magnitude of $D_r$ is markedly higher in the aluminum case, with the peak displacement being four times greater than that observed in the composite wing.

\begin{figure}[t!]
    \centering
    \includegraphics[width = 0.95\textwidth]{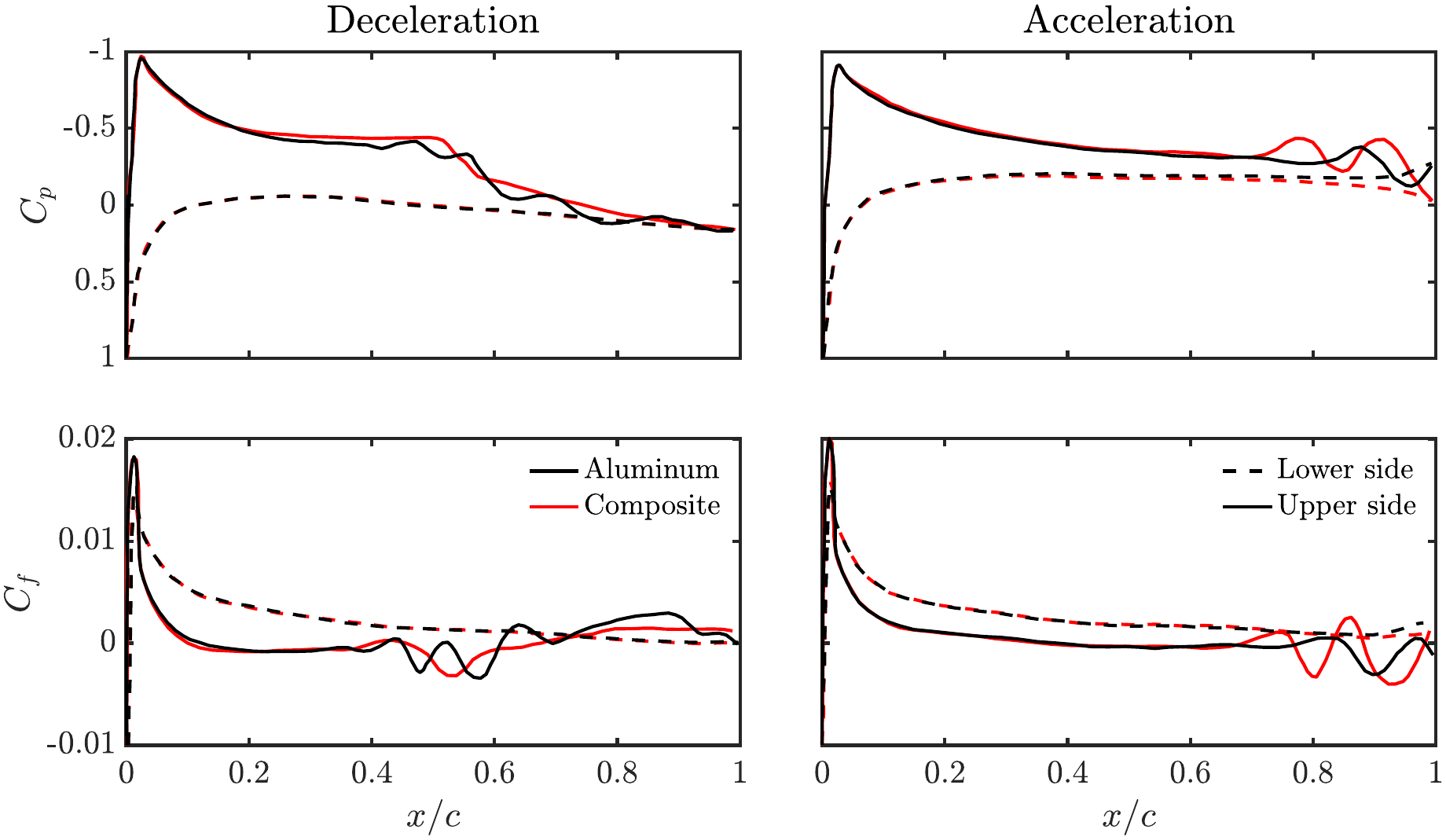}
    \caption{Phase-averaged pressure $C_p$ (upper) and skin-friction $C_f$ coefficients (lower) for both flexible and rigid cases during deceleration ($\phi = 200^\circ$) and acceleration ($\phi = 325^\circ$). Solid lines represent the suction side, while dashed lines denote the pressure side. Note that only 10 gust cycles are employed in this case to perform the phase averaging process to provide a measure of oscillations observed in the fields.}
    \label{fig: CP_CF_uns_FSI}
\end{figure}

Even being relatively low, the displacement should be sufficient to interact or destabilise the boundary layer as demonstrated by Gowree and Atkin \cite{Gowree2022}. At moderate to low Reynolds numbers, the lift is also highly influenced by the viscous nature of the flow as seen in Figure \ref{fig: Curves NACA0012}, where the laminar separation bubble can increase the lift slightly compared to the inviscid lift, or that at very high $Re$. The effects of the surface displacement, despite being relatively small, are evident in the overall lift as shown in Figure \ref{fig: CL_CD_uns_FSI}. This is a result of strong interaction with the laminar separation bubble, which is displayed in the phase-averaged pressure, $C_p$, and skin friction, $C_f$, coefficients in Figure \ref{fig: CP_CF_uns_FSI} during the acceleration and deceleration phases of the gust. The latter is shown for $\phi = 200^\circ$ which corresponds to the maximum $D_r$ difference in Figure \ref{fig: DISP_uns_FSI}. The major differences are seen on the suction side, where the onset of boundary layer separation occurs at similar chordwise position for both cases. On the composite wing, the laminar separation bubble is shorter and experiences a rapid breakdown, settling down to an attached turbulent boundary layer. In contrast, the aluminum wing experiences a series of vortex shedding downstream of the separation point, as evidenced by the peak in local pressure and skin friction coefficients. This is potentially the main source on unsteadiness in the aerodynamic forces and the larger displacement of the surface.

\begin{figure}[t!]
\begin{center}
    \includegraphics[trim={0 1.9cm 0 0},clip, width = \textwidth]{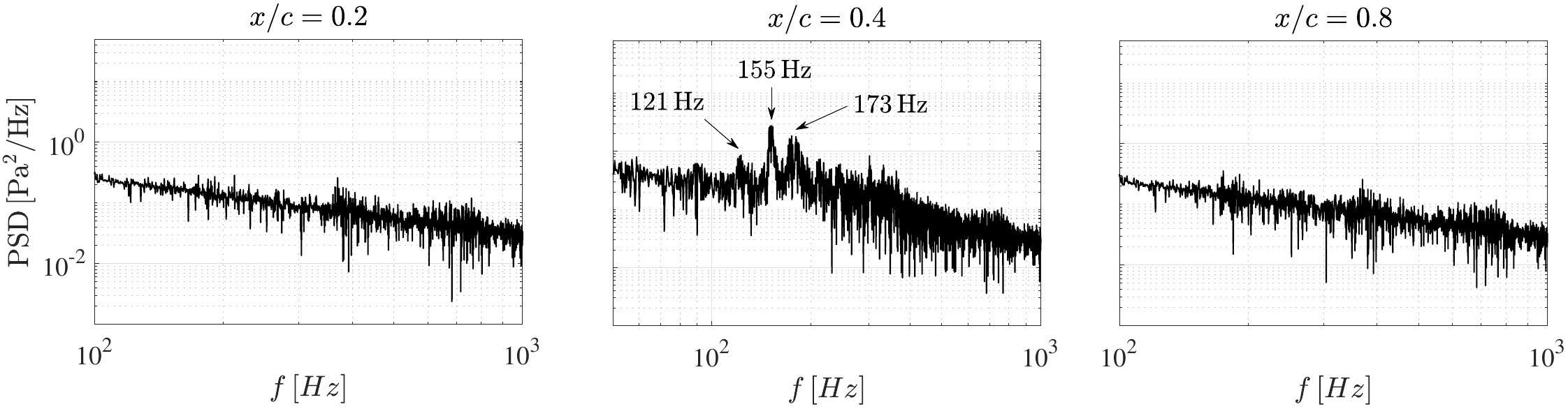}

    \vspace{0.5cm}
    
    \includegraphics[trim={0 0 0 0cm},clip, width = \textwidth]{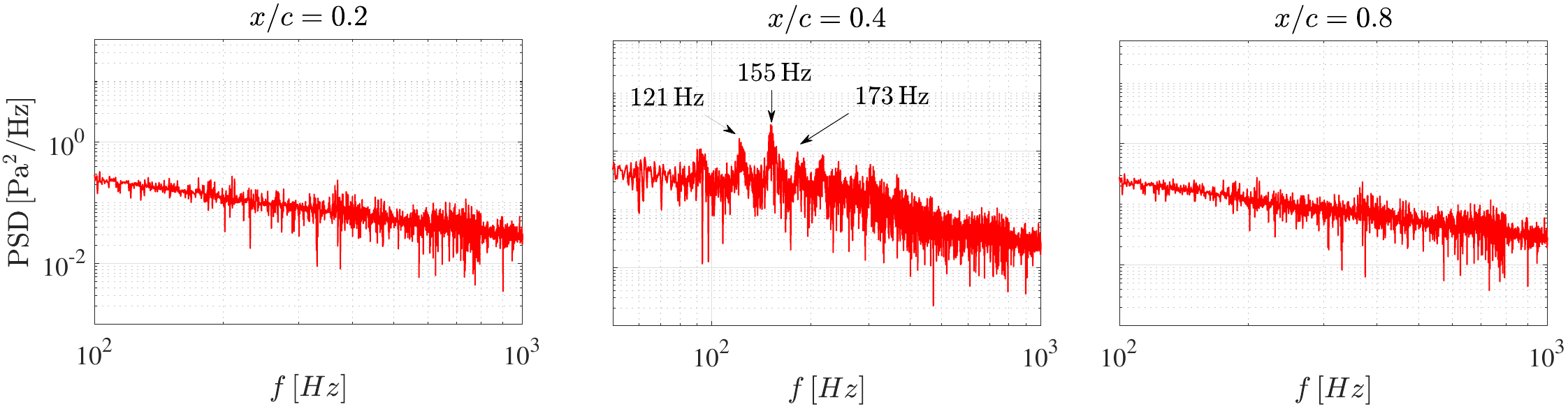}
    \caption{Power spectra of the local streamwise velocity $V_x$ temporal signal, measured at three distinct streamwise locations within the boundary layer. Comparisons are made between the aluminum (black) and composite (red) wing cases.}
    \label{fig: FFT_BL}
\end{center}
\end{figure}

To complement the analysis, the FFT of the time signals of the local streamwise velocity at different chord locations is shown in Figure \ref{fig: FFT_BL}. There is no major difference in the spectra between the two cases at $x/c=0.2$ and $x/c=0.8$, but some interesting behaviours can be observed within the upstream half of separation bubble at, $x/c\approx 0.4$. In particular, there are peaks at $91\,\rm{Hz}$, $121\,\rm{Hz}$, $155\,\rm{Hz}$ and their harmonics in both wings. Interestingly, the main difference between the two spectra at $x/c=0.4$ is observed at $f\approx 173\,\rm{Hz}$, where there is a peak in the aluminum wing spectra, but this mode is less developed on the composite wing. 
It seems that the dynamics of the bubble induces a response in the wing, which is not damped by the aluminum wing because it vibrates at the same frequency. Conversely, the composite wing is capable of dampening part of the oscillations caused by the unsteadiness of the boundary layer, leading to stabilization. This is in-line with the small changes in $C_L$ within $220^\circ<\phi<320^\circ$ and a rapid drop experienced by the composite wing, shown in Figure \ref{fig: CL_CD_uns_FSI}.

\begin{figure}[t!]
\begin{center}
    \includegraphics[width = 0.44\textwidth]{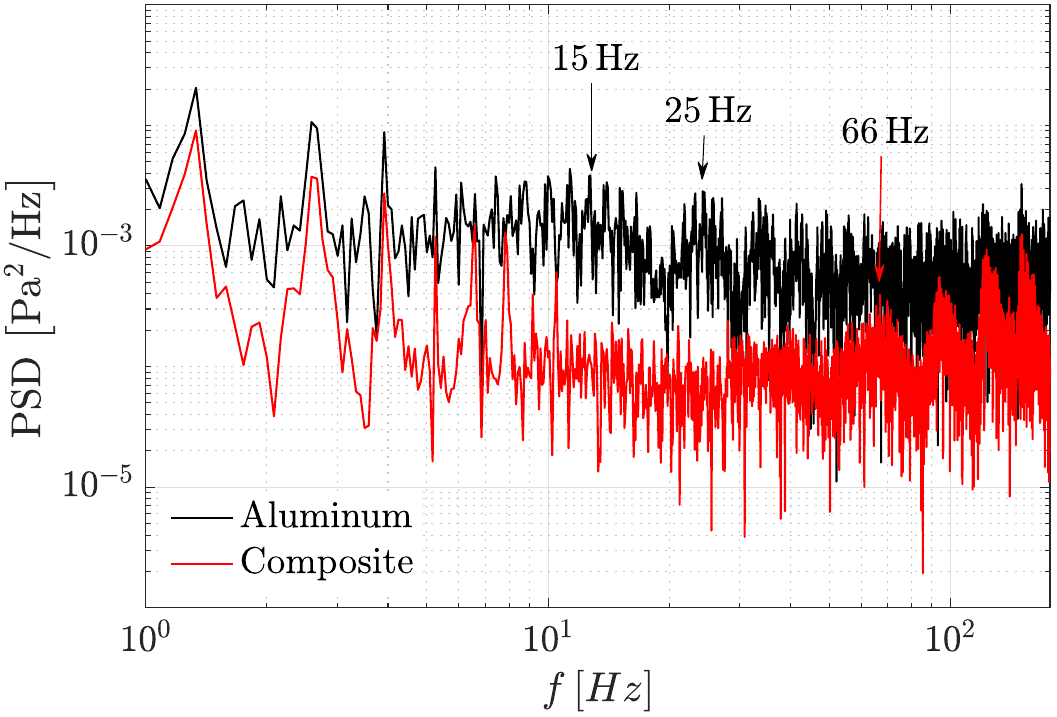}
    \hspace{0.07\textwidth}
    \includegraphics[width = 0.44\textwidth]{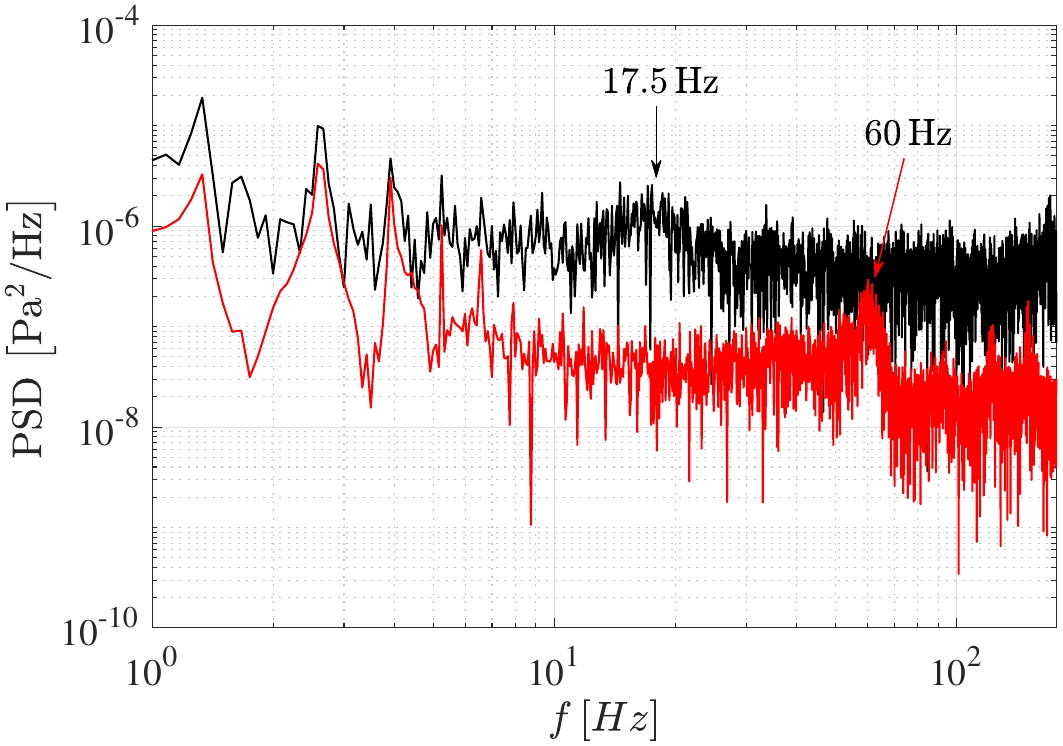}
    \caption{Power spectra of the lift coefficient $C_L$ (left) and maximum relative displacement $D_r$ (right) temporal signals for the two wing materials.}
    \label{fig: fft_cl_dr}
\end{center}
\end{figure}

Finally, Figure \ref{fig: fft_cl_dr} presents the power spectra of the temporal signals for lift coefficient, $C_L$, and maximum relative displacement, $D_r$, for two distinct wing materials. An examination of these spectra can reveal intriguing characteristics associated with the aerodynamic and structural behavior of the wings. For the aluminum wing, the power spectra of $C_L$ distinctly showcases two dominant peaks, located at 15 Hz and 25 Hz. Interestingly, the structural mode of this wing manifests a frequency that resides between these two dominant peaks, specifically at 17.5 Hz. This observation suggests a potential decoupling between the aerodynamic and structural dynamics, and thus the structural vibration is unable to force earlier transition leading to larger unsteadiness from the bubble which favoured the pronounced oscillations observed for the aluminum wing.

Conversely, the composite wing exhibits a more harmonized behavior between its aerodynamic and structural modes. Both modes predominantly centered around a frequency range of 60 to 66 Hz. This alignment in frequencies provides insights into the inherent synergy between the aerodynamic forces and the structural dynamics of the composite wing. The flexibility introduced in the wing results in a structural mode with a higher frequency that appears to lock-in with the aerodynamic forces, leading to a synchronized behavior driven by a common frequency. This might be the reason behind the damping observed in the temporal evolution of the aerodynamic forces and it is in agreement with literature results by Murayama \textit{et al.} \cite{Murayama} for avian feathers. Furthermore, the frequency peaks observed beyond 66 and 60 Hz for the composite wing align closely with the peaks observed in Figure \ref{fig: FFT_BL} for the velocity signals around 121 and 155 Hz in the composite-wing case. This correlation suggests that the aerodynamic and structural behaviors observed are potentially more influenced by the dynamics within the boundary layer for the composite wing.

In contrast, the aluminum wing, characterized by distinct frequencies for the aerodynamic and structural modes, may undergo amplified vibrations and oscillations. This amplification could be ascribed to potential interference and interaction between these two modes, each operating at its unique frequency. Such interactions could result in complex temporal evolutions in both structural and aerodynamic fields, potentially resulting in the observed larger oscillation amplitudes. This is akin to the buffeting phenomenon reported on traditional wings or rotors.

\section{Discussion and conclusion}

This study provides valuable insights into the aerodynamic behaviour of a NACA 0012 aerofoil under both steady and unsteady flow conditions. Drawing inspiration from bird feathers, flexibility of the surface was introduced to potentially control unsteady flow perturbations and alleviate the oscillating aerodynamic forces while flying through gusts.

In the rigid wing analysis, the numerical simulations were validated against an existing experimental dataset, and highlighted the contrasting effects of acceleration and deceleration profiles on the boundary layer under surging flow conditions. While acceleration acted as a favourable pressure gradient and stabilised the boundary layer, deceleration introduced destabilising effects. Interestingly, the dynamics of the Reynolds number were overshadowed by pressure gradient effects in the dynamic case. In regions of the highest Reynolds number, an early onset of separation would be anticipated. However, due to the proximity of this region to the peak acceleration phase, the separation location was observed to be the most downstream.

When the surface was allowed to oscillate and deform during the FSI simulation, the highly unsteady flow due to the presence of laminar separation led to the excitation of the surface, even under steady free stream conditions. Using a composite wing led to a rapid damping in oscillations and a subsequent alleviation of the overall aerodynamic load. In the presence of the gust, the composite wing also exhibited a more rapid recovery and a lower maximum amplitude of deflection. Furthermore, in regions where the pressure gradient effects of the unsteady free stream were most destabilising (during the deceleration phase around $\phi = 200^\circ$), the aluminum wing exhibited large amplitude oscillations. These oscillations were rapidly damped upon introducing composite materials to the wing surface. Furthermore, the composite wing revealed reduced structural displacements compared to its aluminum equivalent.

This behaviour can be attributed to the capability of the composite wing to interact with the dynamics of the laminar separation bubble, fostering a more attached and stable turbulent boundary layer. Consequently, this minimises the unsteadiness of the flow arising from the large vortices typically shed in the laminar separation region. To the authors' knowledge, this kind of behaviour has not been confirmed experimentally on live birds. However, it is reasonable to speculate that in nature, this mechanism would prevent unwanted wing flapping, keeping the strain on the muscles as low as possible to reduce energy expenditure during operation in a perturbed environment.  

\section*{Acknowledgements}
The work was supported by the internal fund of the Department of Aerodynamics Energy and Propulsion at ISAE-SUPAERO and we would like to acknowledge the assistance of the technical team during the wind tunnel experimental campaign.   


 \bibliographystyle{elsarticle-num} 
 \bibliography{main}

\begin{thebibliography}{10}
\expandafter\ifx\csname url\endcsname\relax
  \def\url#1{\texttt{#1}}\fi
\expandafter\ifx\csname urlprefix\endcsname\relax\def\urlprefix{URL }\fi
\expandafter\ifx\csname href\endcsname\relax
  \def\href#1#2{#2} \def\path#1{#1}\fi

\bibitem{hassanalian2017}
M.~Hassanalian, A.~Abdelkefi, Classifications, applications, and design challenges of drones: A review, Progress in Aerospace sciences 91 (2017) 99--131.

\bibitem{jimenez2019}
J.~Jim{\'e}nez~L{\'o}pez, M.~Mulero-P{\'a}zm{\'a}ny, Drones for conservation in protected areas: present and future, Drones 3~(1) (2019) 10.

\bibitem{adkins2019}
K.~A. Adkins, Urban flow and small unmanned aerial system operations in the built environment, International Journal of Aviation, Aeronautics, and Aerospace 6~(1) (2019) 10.

\bibitem{Watkins2006}
S.~Watkins, J.~Milbank, B.~J. Loxton, W.~H. Melbourne, Atmospheric winds and their implications for microair vehicles, AIAA Journal 44~(11) (2006) 2591--2600.
\newblock \href {https://doi.org/10.2514/1.22670} {\path{doi:10.2514/1.22670}}.

\bibitem{Berry2012}
A.~J. Berry, J.~Howitt, D.-W. Gu, I.~Postlethwaite, A continuous local motion planning framework for unmanned vehicles in complex environments, Journal of Intelligent {\&} Robotic Systems 66~(4) (2012) 477--494.
\newblock \href {https://doi.org/10.1007/s10846-011-9633-x} {\path{doi:10.1007/s10846-011-9633-x}}.

\bibitem{Jones2022}
A.~R. Jones, O.~Cetiner, M.~J. Smith, Physics and modeling of large flow disturbances: Discrete gust encounters for modern air vehicles, Annual Review of Fluid Mechanics 54~(1) (2022) 469--493.
\newblock \href {https://doi.org/10.1146/annurev-fluid-031621-085520} {\path{doi:10.1146/annurev-fluid-031621-085520}}.

\bibitem{Jaroslawski2023a}
T.~Jaroslawski, M.~Forte, O.~Vermeersch, J.-M. Moschetta, E.~R. Gowree, Disturbance growth in a laminar separation bubble subjected to free-stream turbulence, Journal of Fluid Mechanics 956 (2023) A33.
\newblock \href {https://doi.org/10.1017/jfm.2023.23} {\path{doi:10.1017/jfm.2023.23}}.

\bibitem{Jaroslawski2023b}
T.~Jaroslawski, M.~Forte, J.-M. Moschetta, E.~R. Gowree, Boundary layer forcing on a rotating wing at low reynolds numbers, Experiments in Fluids 64(58) (2023).
\newblock \href {https://doi.org/10.1017/jfm.2023.23} {\path{doi:10.1017/jfm.2023.23}}.

\bibitem{storer2017}
L.~N. Storer, P.~D. Williams, M.~M. Joshi, Global response of clear-air turbulence to climate change, Geophysical Research Letters 44~(19) (2017) 9976--9984.
\newblock \href {https://doi.org/https://doi.org/10.1002/2017GL074618} {\path{doi:https://doi.org/10.1002/2017GL074618}}.

\bibitem{sachs2013}
G.~Sachs, J.~Traugott, A.~Nesterova, F.~Bonadonna, Experimental verification of dynamic soaring in albatrosses, Journal of Experimental Biology 216~(22) (2013) 4222--4232.

\bibitem{Gowree2018}
E.~R. Gowree, C.~Jagadeesh, E.~Talboys, C.~Lagemann, C.~Brücker, Vortices enable the complex aerobatics of peregrine falcons, Communications Biology 1~(1) (2018) 27.
\newblock \href {https://doi.org/10.1038/s42003-018-0029-3} {\path{doi:10.1038/s42003-018-0029-3}}.

\bibitem{shyy_lian_tang_viieru_liu_2007}
W.~Shyy, Y.~Lian, J.~Tang, D.~Viieru, H.~Liu, Aerodynamics of Low Reynolds Number Flyers, Cambridge Aerospace Series, Cambridge University Press, 2007.
\newblock \href {https://doi.org/10.1017/CBO9780511551154} {\path{doi:10.1017/CBO9780511551154}}.

\bibitem{abbasi2019bio}
S.~Abbasi, A.~Mahmood, Bio-inspired gust mitigation system for a flapping wing {UAV}: modeling and simulation, Journal of the Brazilian Society of Mechanical Sciences and Engineering 41~(11) (2019) 524.

\bibitem{Cheney}
J.~A. Cheney, J.~P.~J. Stevenson, N.~E. Durston, J.~Song, J.~R. Usherwood, R.~J. Bomphrey, S.~P. Windsor, Bird wings act as a suspension system that rejects gusts, Proceedings of the Royal Society B: Biological Sciences 287~(1937) (2020) 20201748.
\newblock \href {https://doi.org/10.1098/rspb.2020.1748} {\path{doi:10.1098/rspb.2020.1748}}.

\bibitem{Schluter}
J.~Schluter, Lift Enhancement at Low Reynolds Numbers Using Pop-Up Feathers, AIAA, 2012, Ch. 4195.
\newblock \href {https://doi.org/10.2514/6.2009-4195} {\path{doi:10.2514/6.2009-4195}}.

\bibitem{Clark}
C.~J. Clark, D.~O. Elias, M.~B. Girard, R.~O. Prum, {Structural resonance and mode of flutter of hummingbird tail feathers}, Journal of Experimental Biology 216~(18) (2013) 3404--3413.

\bibitem{Murayama}
Y.~Murayama, T.~Nakata, H.~Liu, Flexible flaps inspired by avian feathers can enhance aerodynamic robustness in low reynolds number airfoils, Frontiers in Bioengineering and Biotechnology 9 (2021).
\newblock \href {https://doi.org/10.3389/fbioe.2021.612182} {\path{doi:10.3389/fbioe.2021.612182}}.

\bibitem{zhang2018}
S.~Zhang, Z.~Wang, Y.~Wu, Y.~Yu, Flight dynamic coupling analysis of a bio-inspired elastic-wing aircraft, The Aeronautical Journal 122~(1250) (2018) 572–597.
\newblock \href {https://doi.org/10.1017/aer.2018.11} {\path{doi:10.1017/aer.2018.11}}.

\bibitem{Kan2020}
Z.~Kan, D.~Li, J.~Xiang, C.~Cheng, Delaying stall of morphing wing by periodic trailing-edge deflection, Chinese Journal of Aeronautics 33~(2) (2020) 493--500.
\newblock \href {https://doi.org/https://doi.org/10.1016/j.cja.2019.09.028} {\path{doi:https://doi.org/10.1016/j.cja.2019.09.028}}.

\bibitem{bil2013}
C.~Bil, K.~Massey, E.~J. Abdullah, Wing morphing control with shape memory alloy actuators, Journal of Intelligent Material Systems and Structures 24~(7) (2013) 879--898.

\bibitem{barbarino2014}
S.~Barbarino, E.~S. Flores, R.~M. Ajaj, I.~Dayyani, M.~I. Friswell, A review on shape memory alloys with applications to morphing aircraft, Smart materials and structures 23~(6) (2014) 063001.

\bibitem{costanza2020}
G.~Costanza, M.~E. Tata, Shape memory alloys for aerospace, recent developments, and new applications: A short review, Materials 13~(8) (2020) 1856.

\bibitem{vos2007}
R.~Vos, R.~De~Breuker, R.~Barrett, P.~Tiso, Morphing wing flight control via postbuckled precompressed piezoelectric actuators, Journal of Aircraft 44~(4) (2007) 1060--1068.

\bibitem{henry2019}
A.~C. Henry, G.~Molinari, J.~R. Rivas-Padilla, A.~F. Arrieta, Smart morphing wing: optimization of distributed piezoelectric actuation, AIAA journal 57~(6) (2019) 2384--2393.

\bibitem{sofla2010}
A.~Sofla, S.~Meguid, K.~Tan, W.~Yeo, Shape morphing of aircraft wing: Status and challenges, Materials \& Design 31~(3) (2010) 1284--1292.

\bibitem{ferrand_gowree_2022}
V.~Ferrand, E.~R. Gowree, Aerodynamic analysis of transitional wings encountering high amplitude streamwise gust, 3AF Association Aéronautique et Astronautique de France, 2022.

\bibitem{kwSST1994}
F.~R. Menter, Two-equation eddy-viscosity turbulence models for engineering applications, AIAA Journal 32~(8) (1994) 1598--1605.
\newblock \href {https://doi.org/10.2514/3.12149} {\path{doi:10.2514/3.12149}}.

\bibitem{Langtry2009}
R.~B. Langtry, F.~R. Menter, {Correlation-Based Transition Modeling for Unstructured Parallelized Computational Fluid Dynamics Codes}, AIAA Journal 47~(12) (2009) 2894--2906.
\newblock \href {https://doi.org/10.2514/1.42362} {\path{doi:10.2514/1.42362}}.

\bibitem{isaacs1945}
R.~Isaacs, Airfoil theory for flows of variable velocity, Journal of the Aeronautical Sciences 12~(1) (1945) 113--117.

\bibitem{greenberg1947}
J.~M. Greenberg, Airfoil in sinusoidal motion in a pulsating stream, Tech. Rep. NACA-TN-1326, National Advisory Committee for Aeronautics, accession Number: 93R11406 (June 1947).

\bibitem{strangfeld2016}
C.~Strangfeld, H.~M{\"u}ller-Vahl, C.~N. Nayeri, C.~O. Paschereit, D.~Greenblatt, Airfoil in a high amplitude oscillating stream, Journal of Fluid Mechanics 793 (2016) 79--108.

\bibitem{van1994}
B.~G. van~der Wall, J.~G. Leishman, On the influence of time-varying flow velocity on unsteady aerodynamics, Journal of the American Helicopter Society 39~(4) (1994) 25--36.

\bibitem{van1992}
B.~G. Van~der Wall, The influence of variable flow velocity on unsteady airfoil behavior, Ph.D. thesis, University of Maryland, USA (1992).

\bibitem{Tani1964}
I.~Tani, Low-speed flows involving bubble separations, Progress in Aerospace Sciences - PROG AEROSP SCI 5 (1964) 70--103.
\newblock \href {https://doi.org/10.1016/0376-0421(64)90004-1} {\path{doi:10.1016/0376-0421(64)90004-1}}.

\bibitem{Greenblatt2023}
D.~Greenblatt, H.~M\"uller-Vahl, C.~Strangfeld, Laminar separation bubble bursting in a surging stream, Phys. Rev. Fluids 8 (2023) L012102.
\newblock \href {https://doi.org/10.1103/PhysRevFluids.8.L012102} {\path{doi:10.1103/PhysRevFluids.8.L012102}}.

\bibitem{Gowree2022}
E.~R. Gowree, C.~J. Atkin, Excitation of instabilies in a blasius boundary layer by surface vibration, Journal of Fluids and Structures 114 (2022) 103700.
\newblock \href {https://doi.org/https://doi.org/10.1016/j.jfluidstructs.2022.103700} {\path{doi:https://doi.org/10.1016/j.jfluidstructs.2022.103700}}.

\end{thebibliography}





\end{document}